\documentclass[aps,prb,twocolumn,showpacs,preprintnumbers,amsmath,amssymb,superscriptaddress]{revtex4}

\usepackage{graphicx}
\usepackage{dcolumn}
\usepackage{bm}
\usepackage{color}
\usepackage{epstopdf}

\newcommand{\myRe}{\Re e\;}

\newcommand{\expect}[1]{\langle #1 \rangle}
\newcommand{\e}{\varepsilon}
\newcommand{\be}{\begin{equation}}
\newcommand{\ee}{\end{equation}}
\newcommand{\bea}{\begin{eqnarray}}
\newcommand{\eea}{\end{eqnarray}}
\newcommand{\w}{\omega}
\newcommand{\s}{\sigma}

\newcommand{\down}{\downarrow}

\begin{document}

\title{Theory of a.c. spin current noise and spin conductance through a quantum dot \\ in the Kondo regime I: The equilibrium case}

\author{C. P. Moca}
\affiliation{Department of Theoretical Physics, Institute of
Physics, Budapest University of Technology and Economics, H-1521
Budapest, Hungary}
\affiliation{Department of Physics, University
of Oradea, 410087, Oradea, Romania}

\author{I. Weymann}
\affiliation{Department of Physics, Adam
Mickiewicz University, 61-614 Pozna\'n, Poland}

\author{G. Zarand}
\affiliation{Department of Theoretical Physics, Institute of
Physics, Budapest University of Technology and Economics, H-1521
Budapest, Hungary}
\affiliation{Dahlem Center for Complex Quantum Systems and Fachbereich Physik, Freie Universit\" at Berlin, 14195 Berlin, Germany}

\date{\today}

\begin{abstract}
We analyze the equilibrium  frequency-dependent
spin current noise and spin conductance through a quantum dot in the
local moment regime. Spin current correlations behave markedly
differently from charge correlations. Equilibrium spin
correlations are characterized by two universal scaling functions
in the absence of an external field:
one of them is related to charge correlations, while the
other one describes cross-spin correlations. We characterize these
functions using a combination of perturbative and non-perturbative
methods. We find that at low temperatures spin cross-correlations 
are suppressed at frequencies below the Kondo scale, $T_K$, and  
a dynamical spin accumulation resonance is found at the Kondo energy,
$\omega\sim T_K$. At higher temperatures, $T>T_K$, surprising low-frequency
anomalies  related to overall spin conservation appear in the spin
noise and spin conductance, and the Korringa rate is shown to play a 
distinguished role. The transient spin current response also
displays universal and singular properties. 
\end{abstract}



\maketitle


\section{Introduction}


Nanoelectronic devices are likely to provide our future technology and serve as 
basic tools for storing information, quantum computation~\cite{loss02,HansonNature2008} 
or spin manipulation.~\cite{wolf01,zutic04} 
Due to recent developments in fabrication, it is now possible to produce and measure 
spin accumulation in mesoscopic circuits, or filter the generated spin currents.
\cite{yang_NatPhys08,Folk_PRL09,Folk_Nature09,zhao_NatPhys10} 
One of the next most prominent goals of spintronics is to go 
further towards the microscopic regime,
\cite{Hanson_RMP07,nowack_Science07,press_Nature08} and try to 
measure and manipulate {\em single spins} in quantum dots using spin 
biased circuits. Understanding the structure of spin current noise and response 
through quantum dots is therefore of primary importance. 
Moreover, the interplay of strong interactions and the impact of 
quantum fluctuations of the spin on spin transport are 
also important questions of fundamental interest on their own, and 
quantum dots, being the simplest strongly interacting systems, 
also play a prominent role in this regard: they allow for 
the systematic and controlled experimental and theoretical 
study of strongly interacting states. Although 
not easy to measure,~\cite{Clarke,Heiblum,Deblock03,
OnacPRL2006_2,reulet_PRL08,delattre_NatPhys09}
dynamical noise spectra and response functions 
would allow to gain a clear insight to the structure of interactions.~\cite{BlanterPR2000}

In spite of its obvious importance, however, surprisingly little is known about the 
dynamical spin current response and the spin current noise spectrum of a quantum dot. 
Spin current shot noise in the sequential tunneling regime
has been theoretically studied in Ref.~[\onlinecite{SauretPRL2004}], and later 
in the co-tunneling regime by Kindermann.~\cite{KindermannPRB2005}
However, these calculations focussed almost exclusively on  
d.c. properties, and have avoided the strong coupling 
(Kondo) regime,~\cite{Hewson_Book} which is much more difficult to reach 
and understand theoretically.

In the present paper we make an attempt to characterize the 
equilibrium spin noise spectrum and the dynamical spin response of a quantum dot. 
Here we focus on the local moment regime,~\cite{Kouwenhoven_QDsReview}
where charge fluctuations can be neglected, and the dot can simply be
described in terms of a local spin operator, $\mathbf{S}$ ($S=1/2$), 
coupled to the left and right lead electrons through an exchange coupling, $j$.  
We shall study how time dependent spin polarized currents can be 
injected through the dot at various temperatures, and how these 
spin polarized currents fluctuate in time.
To obtain a coherent and clear picture, we combine 
various numerical and analytical methods such as numerical renormalization group (NRG), 
perturbative and renormalization group calculations, and a master equation approach.

Though some of the results presented here are also valid in non-equilibrium, 
in this paper we focus exclusively on the equilibrium case, and leave the
detailed presentation  of the rather technical non-equilibrium  
quantum Langevin calculation to a subsequent publication.\cite{future} 
Nevertheless, even in this equilibrium case, the spin noise and the 
spin conductance display an extremely rich structure; in addition to
the temperature, $T$, and  the Kondo
temperature, $T_K$, below which  the dot spin gets screened, 
new time scales emerge. In the regime, $T\gg T_K$, e.g.,
we find that the Korringa rate (the characteristic spin decay rate), 
\be 
E_K\approx \frac { \pi T }{ \ln^2\left( T/T_K\right)}\;,
\label{eq:Korringa}
\ee   
plays a distinguished role. Furthermore, incorporating the effect of external 
spin relaxation processes  also turns out to be important, and   
introduces an additional new rate, $1/\tau_s$.

\begin{figure}
  \includegraphics[width=0.85\columnwidth]{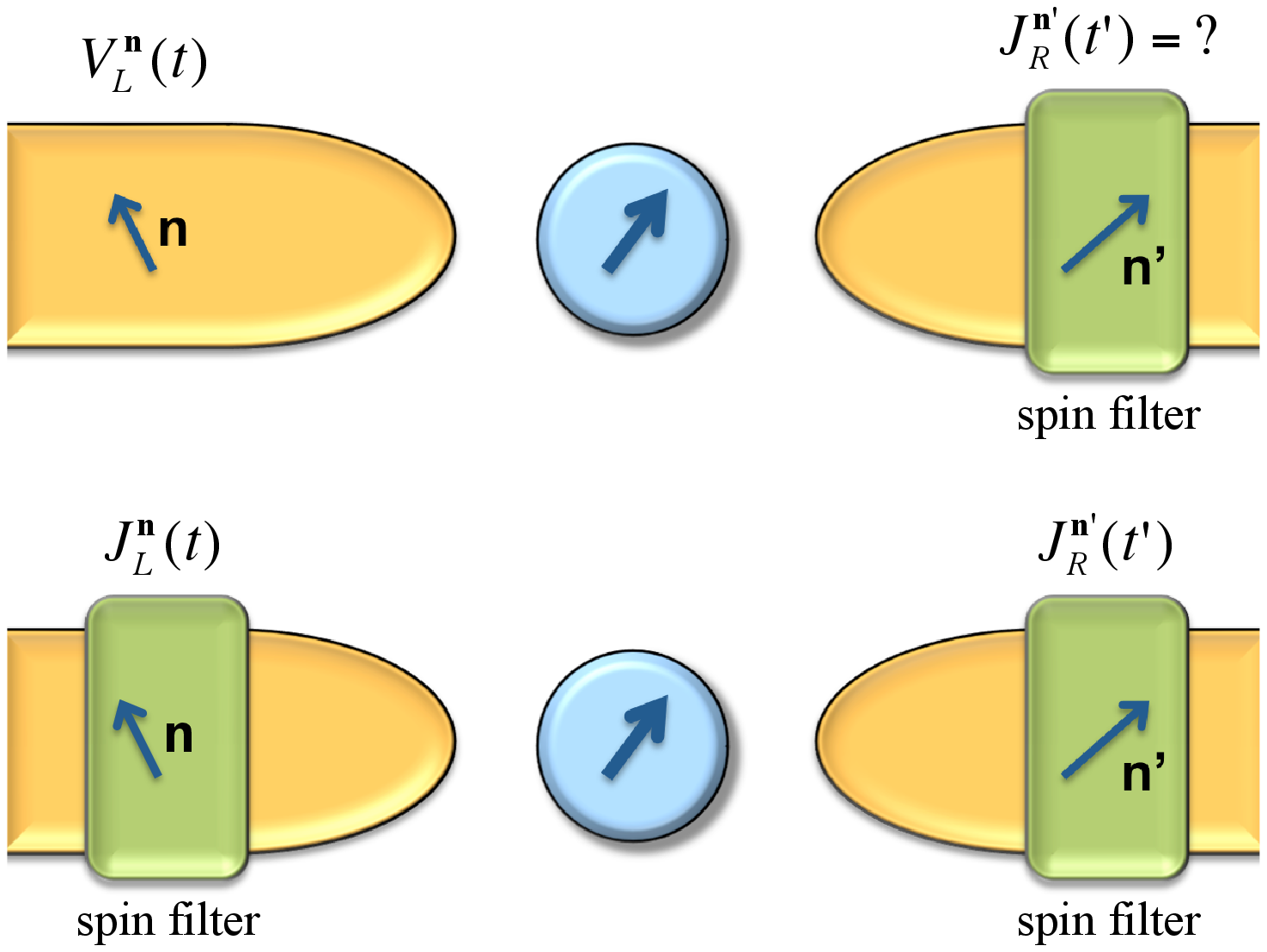}
  \caption{\label{fig:filters}
  (Color online) 
Sketch of the setups to measure 
the spin-dependent left-right conductance, 
$G_{LR}^{\mathbf{n}\mathbf{n'}}(\omega)$, and the noise, 
$S_{LR}^{\mathbf{n}\mathbf{n'}}(\omega)$. 
The arrows indicate the direction of the spin polarization in each lead.
In the upper setup the spin filter on the right side
detects the spin-resolved current at time $t'$, $J_R^{\mathbf n'}(t')$,
induced by applying a spin-dependent voltage to the left lead at time $t$, $V_L^{\mathbf{n}}(t)$.
The lower setup allows for injecting spins and measuring the spin-resolved currents
in both electrodes.}
\end{figure}

In our analysis, we focus on two crucial quantities. 
On the one hand, we study the dynamical spin  conductance 
$G_{rr'}^{\mathbf{n}\mathbf{n'}}(\w)$. This quantity 
characterizes how a current $J_r^{\mathbf{n}}(t)$ of 
carriers  with spins polarized in direction $\mathbf{n}$ is 
generated in the left ($r=L$) or right ($r=R$) lead
by a time dependent chemical potential shift, 
$\delta\mu_{r'}^{\mathbf{n'}}(t) = e V_{r'}^{\mathbf{n'}}(t) $, 
acting in lead $r'$ on carriers polarized along $\mathbf{n'}$ 
(see Fig.~\ref{fig:filters}),
\begin{equation}
  \expect{ J_{r}^{\mathbf{n}}(t)} = \int {\rm d}t'\;
G_{rr'}^{\mathbf{n}\mathbf{n'}} (t-t')\; V_{r'}^{\mathbf{n'}} (t') \,.
\label{eq:linear_response}
\end{equation}
Furthermore, we also investigate time dependent correlations 
of the currents $J_{r}^{\mathbf{n}}(t)$, 
\begin{eqnarray} 
  S_{rr'}^{\mathbf{n}\mathbf{n'}} (t-t') &\equiv& \frac{1}{2}
  \expect{\{ J_{r}^{\mathbf{n}}(t), J_{r'}^{\mathbf{n'}}(t') \}}
\,,
\label{eq:noise_def}
\end{eqnarray}
and determine the corresponding noise spectra.
Of course, in the equilibrium case studied here, 
$G_{rr'}^{\mathbf{n}\mathbf{n'}} $ and 
 $S_{rr'}^{\mathbf{n}\mathbf{n'}}$ are not independent, 
but are related by the fluctuation dissipation
theorem [see Eq.~\eqref{eq:fluctuation_dissipation}].

\begin{figure}[t]
  \includegraphics[width=0.95\columnwidth]{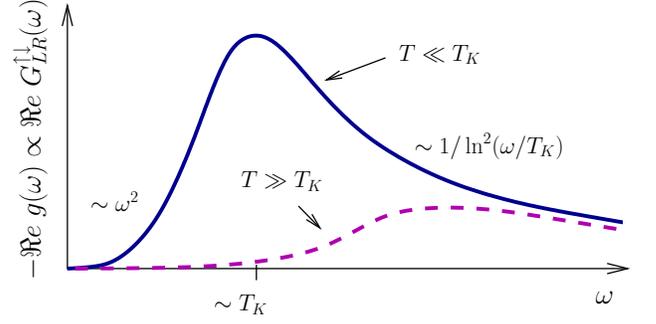}
  \caption{\label{fig:g_simple}
  (Color online)  Sketch of the properties of the real part
of the universal function, 
$-g(\omega,T)\sim G_{LR}^{\uparrow\downarrow}(\w)$, characterizing the 
cross-spin conductance through the dot.}
\end{figure}

One of our main results is, that - in the absence of external spin
relaxation and external magnetic field - both the noise and the conductance take on 
simple, universal forms, and apart from some geometry dependent 
prefactors, are  characterized by just two universal functions. 
The left-right ($r=L$, $r'=R$) conductance, e.g., 
for $\mathbf{n}=\sigma\hat z$ and $\mathbf{n'}=\sigma'\hat z$ reads
\be
G_{LR}^{\sigma\sigma'}(\omega)  =  
\frac{e^2}{h}\;\sin^2(\phi)
\; 
 \left[
\delta_{\sigma\sigma'}
\tilde g  (\omega, T)
+ 
\sigma \sigma'\;
g(\omega, T)  \right]\;,
\label{universalG}
\ee
where only the prefactor depends on the specific geometry of the dot, characterized by the angle
$\phi$ (see Sec.~\ref{section:framework} and Ref.~[\onlinecite{Pustilnik}]),
but the functions $g$ and $\tilde g$ are 
universal functions of $\omega/T_K$ and $T/T_K$. 
The prefactor, $\frac{e^2}{h}$ in Eq.~\eqref{universalG}, denotes 
the universal conductance quantum.~\footnote{Although in this paper we 
use units of $\hbar=k_B=1$, in certain formulas 
we shall restore and display the Planck constant to 
clarify physical dimensions.}
The function $\tilde g$ describes 
charge conductance through the dot, while $g$ 
determines the cross-spin ($\s=\uparrow$, $\s'=\downarrow$)
conductance, $G^{\uparrow\downarrow}_{rr'}(\w)$.
Our main goal is to study the properties of $g$ and those of the corresponding 
noise component in detail. The characteristic features 
of $g$ are shown in Fig.~\ref{fig:g_simple}, which gives a concise summary 
of our most important results. 
As shown in Fig.~\ref{fig:g_simple}, $g$ 
{\em vanishes} in the limit, $\omega\to 0$, and $|g|$ develops a
dip at $\w< T_K$, and a  broad 
resonance at $\omega\sim T_K$ for temperatures $T\ll T_K$. The vanishing 
of the d.c. conductance is a consequence of the fact that spin can only 
 be transferred from the spin-up channel to the spin-down channel by flipping the dot spin, $S$. Thus the amount 
of spin  transfer is limited, and no d.c. spin-cross conductance is possible 
in the absence of external spin relaxation. Temporarily, however, one can 
transfer spin between these two channels at the expense of accumulating spin 
on the dot. This amounts in the appearance of the broad resonance. 

\begin{figure}[b]
 \includegraphics[width=0.95\columnwidth]{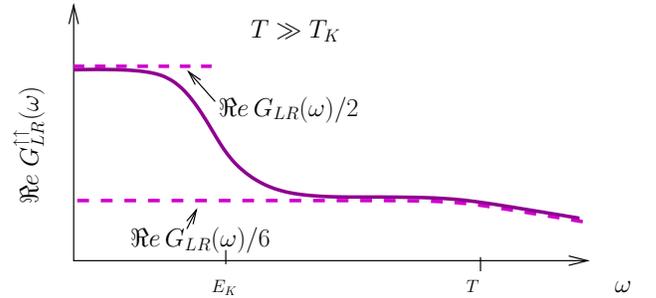}
  \caption{\label{fig:Gupup_simple}
  (Color online) Real part of the conductance 
$ G_{LR}^{\uparrow\uparrow}(\omega)$
through the dot for $T\gg T_K$. The anomaly  below the Korringa relaxation 
rate, $E_K$, is a consequence of correlations between consecutive 
spin-flip processes, and is related to the dip in $|\myRe g(\omega)|$ 
(see Fig.~\ref{fig:g_simple}).}
\end{figure}

The above-mentioned dip in $|\myRe g|$ also survives at temperatures, 
 $T\gg T_K$: There, by the simple argument above, 
 the cross-spin conductance also 
goes to zero as $\omega\to0$, however, this happens only below the 
Korringa rate, $\omega<E_K$ [see Eq.~\eqref{eq:Korringa}], 
where correlations between  consecutive spin-flip events become important. 
As a consequence of    
this, the spin-up -- spin-up conductance, 
$G_{LR}^{\uparrow\uparrow}(\w)$, develops a 
{\em peak} for frequencies $\omega < E_K$.  
The relative size of the peak is universal (see Fig.~\ref{fig:Gupup_simple}), 
and is determined by the
ratio of spin-flip in the high temperature regime vs spin-diagonal scattering processes.

\begin{figure}[t]
  \includegraphics[width=0.95\columnwidth]{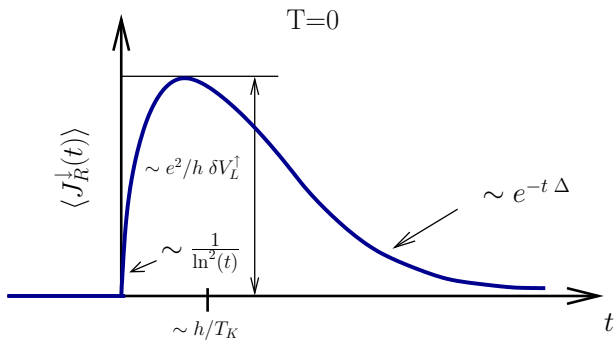}
  \caption{\label{fig:response}
  (Color online) Real time response of the current
$\langle J_R^\downarrow\rangle (t)$, as generated by a 
sudden change of amplitude,  $ \delta V_L^\uparrow$.}
\end{figure}

By just analyzing the analytical structure of the conductance, 
$G^{\sigma\sigma'}_{rr'}(\omega)$, we can also make rather 
general conclusions on the structure of transient response to a sudden potential change at time $t=0$,
$V_{r}^{\sigma}(t)= \delta V_{r}^{\sigma}\; \Theta(t)$, where $\Theta(t)$ denotes the step function.
The linear response is universal, and 
it depends only on $t\, T_K$ and $T/T_K$.
Figure~\ref{fig:response} sketches the structure of the zero-temperature 
response,  $\langle J_R^\downarrow(t)\rangle$ generated by a 
sudden change of  $V_{L}^{\uparrow}$. 
The response is found to be 
logarithmically singular at time $t=0$, 
and at $T=0$ temperature, one observes at a time scale 
$t\approx h /T_K$
an induced current bump of amplitude {$\sim (e^2/h)\;\delta V_{L}^{ \uparrow} $}, followed by an 
exponential decay of the current response. The exponential decay we find 
is somewhat counter-intuitive: the long-time behavior is typically 
associated with Fermi liquid properties,~\cite{Nozieres} 
and therefore one could naively expect an algebraic decay.
However, the exponential decay found follows  precisely
from Fermi liquid theory, which implies certain analytical properties for 
the response functions.

As mentioned before, the properties of the equilibrium noise spectrum are related to those 
of the conductance. \cite{Mikhail} Thus the noise has a structure similar to 
Eq.~\eqref{universalG}, and can also 
be described in terms of just two universal 
functions, $\tilde s(\omega)$ and $s (\omega)$,
describing charge  and cross-spin correlations, respectively.
The left-right noise components
$S^{\sigma\sigma'}_{LR}(\w)$, e.g., can be expressed as 
\begin{equation}
  S_{LR}^{\sigma\sigma'}(\w) = -\frac {e^2} {h} \;T_K\;
  \sin^2(\phi) \; \left[ \delta_{\sigma\sigma'} \tilde s  (\omega, T) +
  {\sigma\sigma'}\;s(\omega, T)  \right] \;,
\end{equation}
with the two universal functions $\tilde s$ and $s$
displaying markedly different behavior.
At $T\to 0$, e.g.,
$\tilde s(\omega)\sim |\omega|/T_K$,
while spin cross-correlations scale as $s(\omega)\sim |\omega|^3/T_K^3$. 

\begin{figure}[t]
\includegraphics[width=0.95\columnwidth]{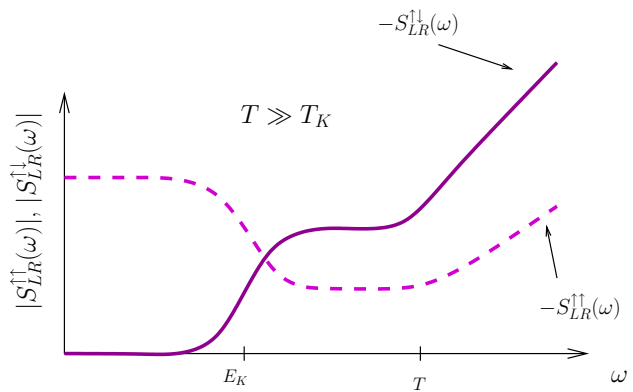}
  \caption{\label{fig:s_simple}  
  (Color online)  Structure of the spin noise 
components $S_{LR}^{\uparrow\uparrow}(\w)$ and 
$S_{LR}^{\uparrow\downarrow}(\w)$ for $T\gg T_K$.}
\end{figure}

This difference between $s$ and $\tilde s$ carries over to finite
temperatures, where $\tilde s$ remains finite in the d.c. 
limit,  while the spin-dependent  component $s$ of the noise
always scales to zero as $\omega\to0$. 
In particular, at temperatures $T\gg T_K$
the charge component of the noise
is similar to what is found for an ordinary tunnel 
junction,~\cite{CallenWelton,Glattli}
\be 
{\tilde  s(\omega,{T\gg T_K}) } 
\approx \frac{3\pi^2}{16} \;  j^2(\w,T)\; 
\frac{\omega}{T_K} \; 
\coth \left(\frac{\w}{2T}\right)\;,
\ee
and remains proportional to the temperature, $T$, for $\w\to0$. 
Here the only nontrivial physics is carried by 
 the prefactor $j^2(\w,T)\approx 1/\ln^2(\max\{|\w|,T\}/T_K)$, 
which  accounts for the 
Kondo-renormalized amplitude of the current, mediated by the 
 exchange processes.
Similar to $g$, however, the spin-dependent part $s$ 
has a non-trivial structure, and 
exhibits a dip at frequencies below the Korringa rate, 
$\w<E_K$, where correlations between consecutive spin-flip 
processes drive the cross-spin noise to zero. This amounts in 
somewhat unexpected 
low frequency  anomalies in the 
noise spectra $S_{rr'}^{\sigma\sigma'}(\w)$, 
as sketched in Fig.~\ref{fig:s_simple} for the particular case of 
the left-right noise components.

The results discussed so far hold in the case, where spin relaxation on the 
dot is induced exclusively by the exchange coupling, $j$, to the lead 
electrons. Then the d.c. 
cross-spin conductance and the shot noise  vanish due to 
spin conservation. External spin relaxation, however, 
slightly modifies the picture above, and 
in its presence we find a small but finite spin shot noise and 
d.c. spin conductance. 
The precise dependence of the noise on the spin relaxation rate, 
$1/\tau_s$,  is rather complex, and is 
analyzed in Sec.~\ref{sec:spin_relaxation} in detail. 
Experimentally, however,  the most relevant regime seems to be
the one, where {$h/\tau_s \ll T,E_K$}. There we find a residual 
cross-spin noise
\be
 S_{LR}^{\uparrow\down}(0) \sim 
\begin{cases} -\frac{e^2}{h} \frac{h}{\tau_s}, & \mbox{if} \;\; T\gg T_K\;, \\ 
- \frac{e^2}{h}  \; T \;\left( \frac{h}{\tau_s T_K}\right)^2,
 & \mbox{if} \;\; T\ll T_K \;.
\end{cases}
\ee
For typical quantum dots, this residual noise and 
the corresponding conductance are thus very small compared to all 
other  features, and therefore  inclusion 
of  $1/\tau_s $ only  slightly modifies the picture obtained 
within the Kondo model.

The paper is organized as follows.
In Sec. II we describe the system's Hamiltonian
and define the spin current operators. We then derive
compact expressions for the spin noise and spin conductance
in terms of universal functions using the Kubo formula.
The zero temperature behavior is discussed
in Sec. III, where we first show the analytical results in the perturbative and Fermi liquid regimes,
and then present NRG results for universal functions obtained in the absence/presence of external magnetic field.
As shown in Section III.B, application of an external magnetic field leads to a more complex behavior and has somewhat similar effect as the external spin relaxation for spin currents 
polarized perpendicular to the applied field.

Section IV is devoted to devoted to the study of the role of a finite temperature.
In the perturbative regime, to determine the $\w$-dependence of universal functions
we use a combination of master equation and perturbation theory,
while in the strong coupling regime we resort to the Fermi liquid arguments.
Transient response of the dot to a sudden switch of a time-dependent
spin-resolved voltage is discussed in Sec. VI, while in Sec. VII
we consider the effect of external spin relaxation on the dynamics of
spin correlations. Finally, the conclusions are given in Sec. VII.


\section{Theoretical framework}
\label{section:framework}

\subsection{Hamiltonian}

Throughout this paper, we focus our attention on the 
local moment regime of a quantum dot, 
where the dot is occupied by a single electron and it can be described in terms of  a spin ${\mathbf S}$, 
coupled to the leads. The Hamiltonian of the system 
consists of a part describing electrons in the leads, $H_{\rm leads}$,
and an interacting part, $H_{\rm int}$, which accounts for
the exchange coupling between the dot and the leads,
$H =  H_{\rm leads} + H_{\rm int}$.
As usual,  electrons in the leads are assumed to be non-interacting, 
and  in the absence of a charge or spin bias,  they
are described by the Hamiltonian,
\begin{equation}
  H_{\rm leads} = \sum_{r,\s}
  \int_{-D}^{D}\e\;
  c^{\dagger}_{r\s}(\e)\, c_{r\s}(\e)\, {\rm d}\e\,,
\end{equation}
with $c^{\dagger}_{r\s}(\e)$ being the creation operator for
a spin-$\s$ electron with energy $\e$ in the left ($r=L$) or
right ($r=R$) electrode and $2D$ being the bandwidth.
The energy $\e$ is measured from the 
chemical potential, and the operators $c_{r\s}(\e)$ satisfy the 
usual anticommutation relations: $\{c^\dagger_{r\s}(\e),
c_{r'\s'}(\e')\} = \delta_{rr'}
\delta_{\s\s'}\delta(\e-\e')$.\footnote{The density of states 
is incorporated in the fields, ${c}_{r\s}(\e)$, to obtain the 
normalization in the main text.}

The interaction part, $H_{\rm int}$, can be most easily constructed in terms 
of the fields, $\psi_{r\s} = \int_{-D}^D
c_{r\s}(\e)d\e$, which destroy electrons of spin $\s$ in
the lead $r$. 
In terms of these fields, the Kondo interaction is given by
\begin{equation}
  H_{\rm int}=  \sum_{r,r'=L,R} \sum_{\sigma,\sigma'} \frac j 2 \;
  v_r v_{r'}\; \mathbf{S} \;\psi_{r\sigma}^\dagger
  \bm{\sigma}_{\sigma\sigma'} \psi_{r'\sigma'}\;, \label{eq:H_int}
\end{equation}
with $j$ a dimensionless coupling related to the Kondo
temperature, $T_K \approx  D\, e^{-1/j}$, and $\bm{\sigma}$ 
the Pauli spin matrices. Here the strength of the 
hybridization of the dot level with the leads
is incorporated in the exchange coupling, $j$, while  
its left-right asymmetry 
appears through the dimensionless parameters, $v_{L/R}$. 
These satisfy the relation $v_L^2  + v_R^2=1$, and are
parametrized by an angle $\phi$ 
as $v_L = \cos (\phi/2)$, $v_R = \sin (\phi/2)$.\footnote{The couplings 
$v_L$ and $v_R$ can be taken to be real.} 
Using this
parametrization the maximal, $T=0$ temperature conductance is given by, 
$G_0 = (2e^2/ h)\; \sin^2 (\phi)$,
where $e^2/h$ is the conductance quantum. 
In the equilibrium case, studied here, the problem can be further 
simplified by noticing that only the 
combination $\Psi\equiv v_L\;\psi_L + v_R \;\psi_R$ occurs
in Eq.~\eqref{eq:H_int}. 
Therefore, rewriting the Hamiltonian in terms of the 'even' ($\Psi$) 
and  'odd' ($ \tilde\Psi$) fields, 
\begin{eqnarray} \label{eq:LR_transformation}
  \Psi_{\s} &\equiv& \cos\left( \frac{\phi}{2}\right) \psi_{L \s} +
  \sin\left ( \frac{\phi}{2}\right ) \psi_{R\s} \nonumber\\
  \tilde{\Psi}_{\s} &\equiv& \sin\left (\frac{\phi}{2}\right ) \psi_{L\s}
  -\cos\left (\frac{\phi}{2}\right ) \psi_{R\s} \,, 
\end{eqnarray}
the interaction part simplifies to
\begin{equation}
  H_{\rm int}=   \frac j 2 \;
   \mathbf{S} \;(\Psi^\dagger
  \bm{\sigma} \Psi)\;, \label{eq:H_int2}
\end{equation} 
and the 'odd' field completely decouples from the dot.

\subsection{Spin current, spin conductance, and spin noise}
\label{sub:Kubo}

To calculate the spin noise and the spin conductance, 
we first need to define the spin current operators.  
Let us first focus on the case, where only the $z$-component is measured, and 
define the operators, 
\begin{equation}
Q_{r}^{\s} =e\;\int_{-D}^D c_{r\s}^{\dagger}(\e) c_{r\s}(\e) d\e ,
\end{equation}
measuring the total charge of the electrons in lead $r$ and spin 
component, $\s$, where $e$ is the electron charge.
The corresponding current operator can then be defined as 
\be 
J_{r}^{\s} \equiv \frac{{\rm d}Q_{r}^{\s}}{{\rm d}t}
\;.
\ee
Notice that with this definition a positive current implies 
a charge flow towards the leads. 
Using the equation of motion we obtain 
\begin{eqnarray}
\label{eq:current_operator}
  J_{r}^{\s} &=&  ie \;\frac{j}{2}
  \sum_{r',\sigma'} v_r v_{r'} {\mathbf S}
  \left(
    \psi_{r\s }^{\dagger} \bm{\sigma}_{\s\s'} \psi_{r'\s'}-
    {\rm h.c}
  \right)\;.
\end{eqnarray}
The charge current is 
then simply expressed as the sum of spin currents, 
$J_r = \sum_\s J_{r}^{\s}$.

At this point, it is useful to also express the current operators 
in terms of the even and odd fields, $\Psi$ and $\tilde \Psi$.
Introducing the so-called   composite fermion
operators,~\cite{CostiPRL94} 
$F_\sigma \equiv (\mathbf{S \,{\bm {\sigma}}}\Psi)_\sigma$,
we can write the current operator as a sum of two 
components, 
\begin{eqnarray}
J_{r}^{\s}& = &I_{r}^{\s}+\tilde I_{r}^{\s}\;,
\nonumber
\\
  I_{r}^{\s} &=&  e\;j \;\gamma_{r}\; i( \Psi_\sigma^\dagger  F_\sigma
- F^\dagger_\sigma \Psi_\sigma ) \;, \nonumber
  \\
  \tilde{I}_{r}^{\sigma} &=&  e\;j \;\tilde \gamma_{r} \; i
  (\tilde {\Psi}_\sigma^\dagger  F_\sigma - 
F^\dagger_\sigma \tilde \Psi_\sigma )\;,
\label{eq:Js}
\end{eqnarray}
with the prefactors defined as
$\gamma_{L/R} = (1 \pm \cos(\phi))/4$, and 
$\tilde \gamma_{L/R} = \pm {\sin(\phi)}/{4}$.

It is instructive to analyze the properties of the current operators, 
Eq.~\eqref{eq:Js}. The components $I_{r}^{\sigma}$ satisfy
\be
I_{r}^{\uparrow} +  I_{r}^{\downarrow} \equiv 0\;,
\label{Jprop}
\ee
implying that the components $I_{r}^{\s}$ do not contribute to the 
charge current, $J_r = \sum_\s \tilde I_{r}^{\s}$.
Furthermore, since the components {  $\tilde I_{r}^{\s}$}, satisfy 
current conservation at the operator level, 
 \be
\tilde I_{L}^{\s} + \tilde I_{R}^{\s} \equiv 0\;,
\ee
there is no charge accumulation on the dot,  and $J_L(t)+J_R(t) = 0$
is satisfied at the operator level at any time $t$. 
The current components, $I_{r}^{\s}$, on the other hand, do not 
satisfy current conservation, $I_{L}^{\s}+I_{R}^{\s}\ne 0$, implying that 
spin accumulation is possible on the dot. These properties 
follow naturally  in the local moment (Kondo) limit, where charge 
fluctuations are completely suppressed, and only spin fluctuations 
are allowed on the dot.

\subsection{Conductance and noise}
\label{sub:conductance}

We shall now analyze the structure of the current 
response of the dot generated by an external  perturbation,
$\delta H = V_{r}^{\sigma}(t) \; Q_{r}^{\sigma}(t)$, 
which shifts the potential of carriers with spin $\s$ in lead $r$. 
In linear response, the current 
is determined by the conductance 
$G_{rr'}^{\s\s'}$  in Eq.~\eqref{eq:linear_response},
 given by the Kubo formula, 
\begin{equation}
  G_{rr'}^{\s\s'}(t,t') = - i \Theta(t-t')
  \expect{[J_{r}^{\s}(t),Q_{r'}^{\s'}(t')]} \,. 
\end{equation}
This expression still contains the charge operators, which are 
non-local operators. 
To eliminate them, we differentiate with respect to 
time $t'$ and use the equation of motion. After Fourier transformation we then 
obtain the following form of the Kubo formula, 
\begin{equation} 
\label{eq:EOM_for_GS}
  i \w\,  
G_{r r'}^{\s \s'}(\w) = (J_r^\s;J_{r'}^{\s'})_\omega - (J_r^\s;J_{r'}^{\s'})_{\omega=0}\;,
\end{equation}
where we introduced a compact notation for the Fourier transform 
of the retarded correlation function of any 
two operators, 
\be 
{\cal G}^R_{AB}(\omega)\;\rightarrow \; (A;B)_\omega\;.
\ee
The second term in Eq.~\eqref{eq:EOM_for_GS} originates from the 
discontinuity  of $G_{r r'}^{\s \s'}(t)$ at $t=0$, which can be shown 
to be real and just equal to  $(J_r^\s;J_{r'}^{\s'})_{\omega=0}$.

Let us now focus on the $SU(2)$ symmetrical case. 
To simplify the expression of $G_{r r'}^{\s \s'}$, 
we can make use of the decomposition of $I_r^\s$, 
Eqs.~\eqref{eq:Js}, and exploit 
the property \eqref{Jprop} as well as the fact that 
all cross-correlations of $I_r^\s$ and $\tilde I_{r'}^{\s'}$ vanish. 
Taking furthermore the spin symmetry into account we find that 
$G_{r r'}^{\s \s'}$ reduces to the 
following form, 
\begin{equation}
\label{eq:conductance_universal}
  { G}_{rr'}^{\sigma\sigma'}(\omega)  =  
{e^2\over h}\;  
\left[ \delta_{\sigma\sigma'} \; {\tilde a}_{rr'} \; \tilde g  (\omega, T) +
  {\sigma\sigma'}\; {a}_{rr'} \; g(\omega, T)  \right]\;.
\end{equation}
Here the information 
on the geometry of the dot is exclusively carried by 
the matrices $\bm{\tilde a}(\phi)$ and $\bm{a} (\phi)$,
\begin{equation}
 \bm{\tilde a} = \left (
\begin{array}{cc}
-\sin^2\phi & \sin^2\phi \\
\sin^2\phi & -\sin^2\phi \\
\end{array}
\right), \hspace{0.3cm} 
\bm{a} = \left (
\begin{array}{cc}
4\cos^2\frac{\phi}{2} & \sin^2\phi \\
\sin^2\phi & 4\sin^2\frac{\phi}{2} \\
\end{array}
\right),\nonumber
\end{equation}
while the  functions $g$ and $\tilde g$ are completely 
independent of the  dot geometry. They are  defined in terms 
of the 'reduced current operators', 
\bea
{\cal I}_\s \equiv i (\Psi^\dagger_{\s}F_\s -\rm{h.c.})\;
\\
{\cal \tilde I}_\s \equiv i (\tilde \Psi^\dagger_{\s}F_\s -\rm{h.c.})\;
\eea
and are given by the following expressions,
\bea
g(\omega) = +\frac{\pi j ^2}{8 i \omega}\left[
({\cal I}_\uparrow;{\cal I}_\uparrow)_\omega - 
({\cal I}_\uparrow;{\cal I}_\uparrow)_{\omega=0}
\right ] \;,
\nonumber
\\
\tilde g(\omega) = - \frac{\pi j ^2}{8 i \omega}\left[
({\cal \tilde I}_\uparrow;{\cal \tilde I}_\uparrow)_\omega - 
({\cal \tilde I}_\uparrow;{\cal \tilde I}_\uparrow)_{\omega=0}
\right ] \;.
\label{eq:gs}
\eea
Here we have chosen a normalization such 
that  for  
the ordinary conductance through the dot we recover the 
well-known expression, 
\be 
G_{LR}(\omega,T) = 2\;
{\frac {e^2}{h} }
\;\sin^2(\phi) \;\tilde 
g(\omega,T)\;, 
\ee
with $\tilde g(\omega,T\to 0)=1$ in the unitary limit. 

Importantly, both $g(\omega,T)$
and $\tilde g(\omega,T)$ are dimensionless functions, and, apart from 
some universal prefactors, they both determine physically 
measurable quantities. By the basic principles 
of the renormalization group,  this immediately 
implies that in the scaling limit, $j\to0$, $D\to \infty$, and 
$T_K$ finite, they must both reduce to functions of the ratios,
$\w/T_K$ and $T/T_K$. This can, of course, be checked explicitly by 
performing perturbation theory in the exchange coupling, $j$.

In the equilibrium case, to which we restricted ourselves here, 
the symmetrized current noise 
[Eq.~\eqref{eq:noise_def}] is simply related to the 
retarded Green's functions, $(J_r^\s;J_{r'}^{\s'})_{\omega} $, and
thus to 
the conductance  by the fluctuation-dissipation theorem, 
\begin{equation}
\label{eq:fluctuation_dissipation}
  S_{rr'}^{\s\s'}(\w) =  - \w \coth \left(\frac{\w}{2T}\right)
  {\Re e} \, G_{rr'}^{\s\s'}(\w) \,.
\end{equation}
Using Eq.~\eqref{eq:conductance_universal},
we then immediately find
\begin{equation}
  S_{rr'}^{\sigma\sigma'}(\w) = -
{ \frac {e^2} {h} }
\;T_K\;
   \left[ 
\delta_{\sigma\sigma'} \tilde a_{rr'}\tilde s  (\omega, T) +
  {\sigma\sigma'}\;a_{rr'}\;s(\omega, T)  \right] \;,
  \nonumber
\end{equation}
where $s$ and $\tilde s$ are two real dimensionless universal functions 
characterizing the symmetrized equilibrium noise, 
\begin{equation}
\label{eq:FDT}
  \left (\begin{array}{c}
  s(\omega,T)\\
  \tilde s(\omega, T)
  \end{array} \right ) = \left (
  \begin{array}{c}
  \Re e\;g(\omega,T)\\
   \Re e\;\tilde g(\omega, T)
  \end{array}\right )\;
  \frac \omega {T_K}\; {\rm coth}\left(\frac \omega {2T}\right
  )\,.
\end{equation}
The structures of the universal functions $g$, $\tilde g$, 
$s$, and $\tilde s$ shall be thoroughly discussed in the 
following Sections.

\subsection{Non-collinearity}
\label{sec:Non-collinearity}

The discussion above can readily be generalized to the case, where the 
polarizations of the injected and detected currents are arbitrary, 
but the system still exhibits SU(2) symmetry. 
In this case we first define the charge of carriers 
polarized along $\mathbf{n}$ as 
\begin{equation}
Q_{r}^{{\mathbf n}} \equiv e\, \sum_{\s,\s'}\int_{-D}^D c_{r\s}^{\dagger}(\e)
\frac 1 2 (\mathbf{1}+ \mathbf{n}\,\boldsymbol{\s})_{\s\s'}  
 c_{r\s'}(\e) d\e \;.  
\end{equation}
The corresponding current operators, $J_{r}^{\mathbf{n}}= \frac{\rm
d}{{\rm d} t} Q_{r}^{\mathbf{n}} $ 
can be expressed as 
\be 
J_{r}^{\mathbf{n}} =  I_{r}^{\mathbf{n}}+ \tilde{I}_{r}^{\mathbf{n}} \;,
\nonumber 
\ee
the even and odd current components being defined as
\begin{eqnarray}
  I_{r}^{\mathbf{n}} &=&  e\;j \;\gamma_{r}\; i
( \Psi^\dagger\, {\rm P}_{\mathbf n}\, F
- F^\dagger\,{\rm P}_{\mathbf n}\, \Psi ) \;, \nonumber
 \\
  \tilde{I}_{r}^{\mathbf{n}} &=&  e\;j \;\tilde \gamma_{r} \; i
  (\tilde {\Psi}^\dagger \,{\rm P}_{\mathbf n}\,
  F - F^\dagger \,{\rm P}_{\mathbf n}\,\tilde \Psi )\;,
\end{eqnarray}
with the projector ${\rm P}_{\mathbf n}$ given by 
\begin{equation}
{\rm P}_{\rm n} = (\mathbf{1} + \mathbf{ n}\, \boldsymbol{\s})/ 2\;.
\end{equation}

Apart from the above changes of definition, the calculations 
presented in Subsection~\ref{sub:conductance}
 trivially generalize to  this case, and 
the final expression of 
the conductance assumes the following simple form, 
\begin{eqnarray}
G_{rr'}^{ \mathbf{n}\mathbf{n'}}(\omega) & =  &
{ {e^2\over h} }
 \,
 \Bigl [ \tilde a_{rr'}
{\mathbf{1} + \mathbf{n}\,\mathbf{n'} \over 2}\;
 \tilde g  (\omega, T) 
\nonumber\\
&+ &  a_{rr'}   \; {\mathbf{n}\,\mathbf{n'}}\;
g(\omega, T)  \Bigr ], 
\end{eqnarray}
with the functions $g$ and $\tilde g$ defined by 
Eqs.~\eqref{eq:gs}.
Through the fluctuation-dissipation theorem, we then obtain 
the following expression for the noise, 
\begin{eqnarray}
S_{rr'}^{ \mathbf{n}\mathbf{n'}}(\omega) & =  &
-{  {e^2\over h}} 
\,T_K\,
 \Bigl [ \tilde a_{rr'}
{\mathbf{1} + \mathbf{n}\,\mathbf{n'} \over 2}\;
 \tilde s  (\omega, T) 
\nonumber\\
&+ &  a_{rr'}   \; {\mathbf{n}\,\mathbf{n'}}\;
s(\omega, T)  \Bigr ], 
\end{eqnarray}
with the functions $s$ and $\tilde s$  
defined in Subsection~\ref{sub:conductance}.
 We emphasize again that the above expressions hold only in the 
presence of spin rotation invariance, and in an external magnetic field
the conductance cannot be parametrized in terms of just two 
universal functions (see Sec.~\ref{sec:NRG}).


\section{The $T=0$ temperature limit}

Before discussing the more complex case of 
finite temperature, let us focus on the much simpler $T=0$ case. 
There we can compute the universal scaling functions $g$ and $s$ 
"numerically exactly" using the machinery of
numerical renormalization group (NRG),~\cite{WilsonRMP75,BullaRMP08}
and can understand all their important features 
relatively easily by combining  perturbative renormalization group 
methods with  Fermi liquid arguments.

\subsection{Analytical considerations}
\label{sec:T=0_analytics}

{\em Perturbation theory.}  The simplest one can do 
to compute the noise is to evaluate the current-current correlation 
functions order by order in $j$. The 0-th order contribution to 
$S_{LR}^{\s\s'}$ is found to be
\be 
 S_{LR}^{\uparrow \uparrow} = \frac 12
  S_{LR}^{\uparrow \downarrow} = -\frac{e^2}{h} \;\sin^2\phi
\;|\w|\;\frac {\pi^2 j^2}{16},
\ee
from which we can extract 
$g$ and $\tilde g$ using Eqs.~\eqref{eq:conductance_universal}
and \eqref{eq:fluctuation_dissipation},
\be
\tilde g = \frac{3\pi^2 j^2}{16} + \dots\;;\phantom{nn}
g = - \frac{\pi^2 j^2}8+\dots\;.
\ee
Evaluating higher order terms to $S_{LR}^{\s\s'}$, one 
obtains logarithmic corrections to $g$ and $\tilde g$. 
The so-called leading logarithmic corrections 
can be summed up by a perturbative renormalization group 
procedure,~\cite{Abrikosov,FowlerZawadowski} which amounts in simply replacing 
$j\to j(\w)=1/\ln(\omega /T_K)$, and gives 
\be
\tilde g \approx \frac{3\pi^2 }{16} \frac{1}{\ln^2(|\omega| /T_K)}
\;;\phantom{nn}
g \approx - \frac{\pi^2}8\frac{1}{\ln^2(|\omega| /T_K)} \;,
\label{g:largeomega}
\ee
for large frequencies, $|\w|\gg T_K$. By the fluctuation dissipation theorem, 
Eq.~\eqref{eq:FDT}, we then find for $|\w|\gg T_K$ 
\be 
\tilde s \approx \frac{3\pi^2 }{16} \,
\frac{{|\w|}/{T_K}}{\ln^2(|\omega |/T_K)}
\;;\phantom{nn}
s \approx - \frac{\pi^2}8 \; \frac{{|\w|}/{T_K}}{\ln^2(|\omega| /T_K)} \;.
\label{s:largeomega}
\ee

{\em Fermi liquid regime.} The above expressions approximate the 
scaling functions only at large frequencies, $|\w|\gg T_K$. In the opposite 
limit, $|\w|\ll T_K$, perturbation theory in $j$ breaks down. However, 
the relevant processes can be captured by a simple Fermi liquid 
approach.~\cite{Nozieres} In this very small frequency limit both 
$g$ and $\tilde g$ are analytical. At $T=0$ temperature the charge conductance is unitary
in the limit, $\w\to0$, and correspondingly,
\be
\tilde g =  1+{\cal O}(\w^2)\;,\phantom{nn}\tilde 
s(\omega/T_K) = \frac{|\w|}{T_K} +\dots\;.
\label{eq:tilde_g_FL}
\ee

In contrast to $\tilde g$, however, 
$G_{rr'}^{\uparrow\downarrow}(\omega\to 0)$ and thus 
$g(\w\to0)$ must vanish. This follows from the 
simple observation that it is impossible to generate a steady spin-down 
current in any of the leads  by injecting only spin-up
electrons. This simply follows from Fermi liquid
theory.~\cite{Nozieres}
According to this latter, at $T=0$ temperature the 
impurity spin is completely screened by the Kondo effect, and 
electrons injected at the Fermi energy 
are only subject to elastic scattering. 
As a consequence, 
a  spin-up electron injected right at the Fermi energy 
conserves its spin. 
Electrons (quasiparticles),
however, interact with each-other through a local interaction at the
impurity site. 
For electrons  injected with  energy $\w$,
 this residual electron-electron interaction generates 
spin-flip processes, and leads to a finite spin cross-conductance. 
By simple phase space arguments, the amplitude of
these latter processes scales with the square of the 
energy of the incoming electron, 
$\sim \omega^2$, and therefore $g(\w)\sim \w^2$.
We thus conclude that 
\bea
g =  - \alpha \frac{\w^2}{T_K^2}\;+ \dots\;,
\label{g:smallomega}\phantom{nn}
s =  - \alpha \frac{|\w|^3}{T_K^3}\;+ \dots\;,
\eea
with $\alpha$ a universal parameter.


\subsection{NRG results}
\label{sec:NRG}

In Section~\ref{sub:Kubo} we related the functions $g$, and $\tilde g$
to the correlation functions of the local operators, 
${\cal I}_\sigma$ and ${\cal \tilde I}_\sigma$, respectively
[see Eq.~\eqref{eq:gs}] .
At $T=0$ temperature, such local correlation 
functions can be 
accurately computed  by NRG calculations,~\cite{WilsonRMP75,BullaRMP08} 
which, as we show below, are indeed in full agreement with 
our previous analytical considerations.

\subsubsection{No external field, $B=0$}

Let us start by sketching how one can perform the NRG 
calculations in the case where we have no external magnetic field, $\bm{B}$.
To compute $\tilde g$, we exploit the fact that correlation functions 
of the operators $ \tilde \Psi^\dagger_{\s}F_\s$
and $ F_\s^\dagger \tilde \Psi_{\s}$
can be factorized onto correlation functions 
of $F^\dagger_\s$ and $\tilde \Psi^\dagger_\s$. The latter
being trivial, after some algebra we then obtain for the real part of 
$\tilde g$, 
\bea
\label{eq:tildeg}
&&\myRe  \tilde g(\omega, T) =
\\
&&\phantom{N}\frac{\pi^2 j^2} 4 \;
\int \frac {d\omega' } {2\omega}\;
  \varrho_F (\omega', T)\; \left [f(\omega' -\omega)-
  f(\omega' + \omega)  \right ]\;,
\nonumber
\eea
with $f(\omega)$ the Fermi function and
$ \varrho_F(\omega, T)\equiv - \frac 1 \pi\;
\Im m\; (F_\uparrow;F_\uparrow^\dagger)_\w$ the spectral function of the 
composite fermion operator. 
The prefactor $\pi^2 j^2/4$ in Eq.~\eqref{eq:tildeg}
can be eliminated by  observing that $\tilde g (0,0)=1$,~\cite{Pustilnik}  
and therefore 
\be 
\frac{\pi^2 j^2} 4\; \varrho_F(0, 0) =1\;. 
\ee
Using this relation, we can also express the real part of
the scaling function $g$ as
\begin{equation}
\myRe  g(\omega, T) = - \frac 1  {2\omega}\;
  \frac{\varrho_{{\cal I}_\uparrow {\cal I}_{\uparrow}}(\omega, T)}{ \varrho_F(0, 0)}
\label{eq:Reg_NRG_expression}
\end{equation}
with $\varrho_{{\cal I}_\uparrow\;{\cal I}_{\uparrow}}(\omega, T)
=- \frac 1 \pi\;
\Im m\; ({\cal I}_\uparrow;{\cal I}_\uparrow^\dagger)_\w$
the spectral function of the operator, 
${\cal I}_\sigma \equiv i(\Psi_\sigma^\dagger  F_\sigma-
F^\dagger_\sigma \Psi_\sigma)$. Computing thus the local 
spectral functions, $\varrho_F$ and $\varrho_{{\cal I}_\uparrow {\cal I}_{\uparrow}}$ 
we can determine the real parts of the functions, $g$ and $\tilde g$. 
From these, we can then directly determine the universal 
spin and charge current noise functions $s$ and $\tilde s$.

\begin{figure}[t]
  \includegraphics[width=0.9\columnwidth]{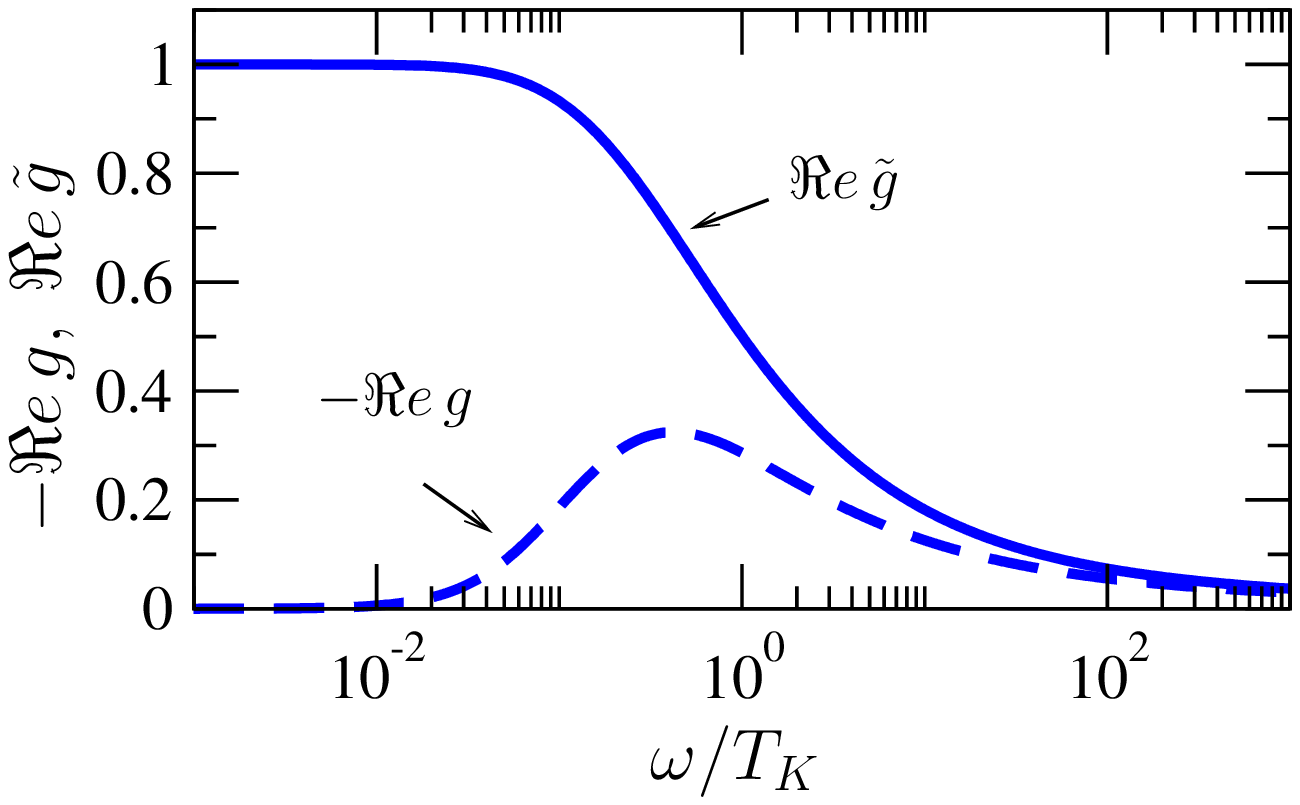}
  \includegraphics[width=0.9\columnwidth]{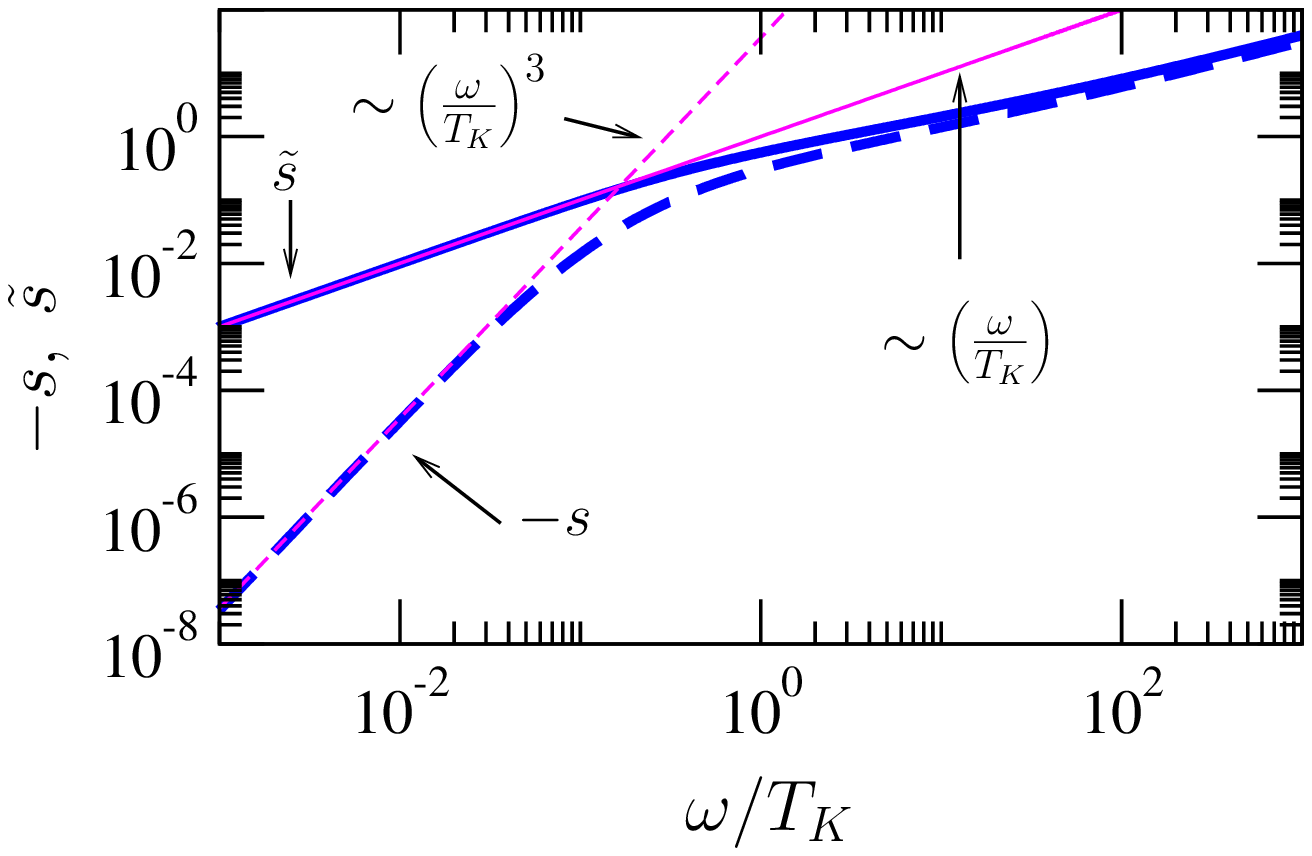}
  \caption{\label{fig:universal_functions}
  (Color online)
  The zero temperature universal functions $g$, $\tilde g$
  for the {a.c.}-conductance (upper panel)
  and for the spin noise $s$ and $\tilde s$ (lower panel)
  computed by NRG.}
\end{figure}

The real parts of $g$ and $\tilde g$ and the corresponding 
scaling functions, $s$ and $\tilde s$, as obtained by NRG
\footnote{In NRG calculations we kept $1024$ states at each iteration and 
assumed $\Lambda=1.8$ and $j=0.18$.}
are shown in Fig.~\ref{fig:universal_functions}. The results presented were 
obtained by the Budapest Flexible-DMNRG code,~\cite{Toth_PRB08,FlexibleDMNRG}
and clearly display all features discussed in the previous subsection.
All functions were plotted as a function of $\omega / T_K$, with 
$T_K$ defined as a half-width
of the function $\myRe \tilde g$. This latter is essentially the a.c. 
charge conductance 
through the dot, studied in Ref.~[\onlinecite{SindelPRL2005}],
and 
displays the expected Kondo resonance, which  appears as a peak 
at frequencies $\w< T_K$. 

\begin{figure*}
  \begin{center}
      \includegraphics[width=1.7\columnwidth, clip]{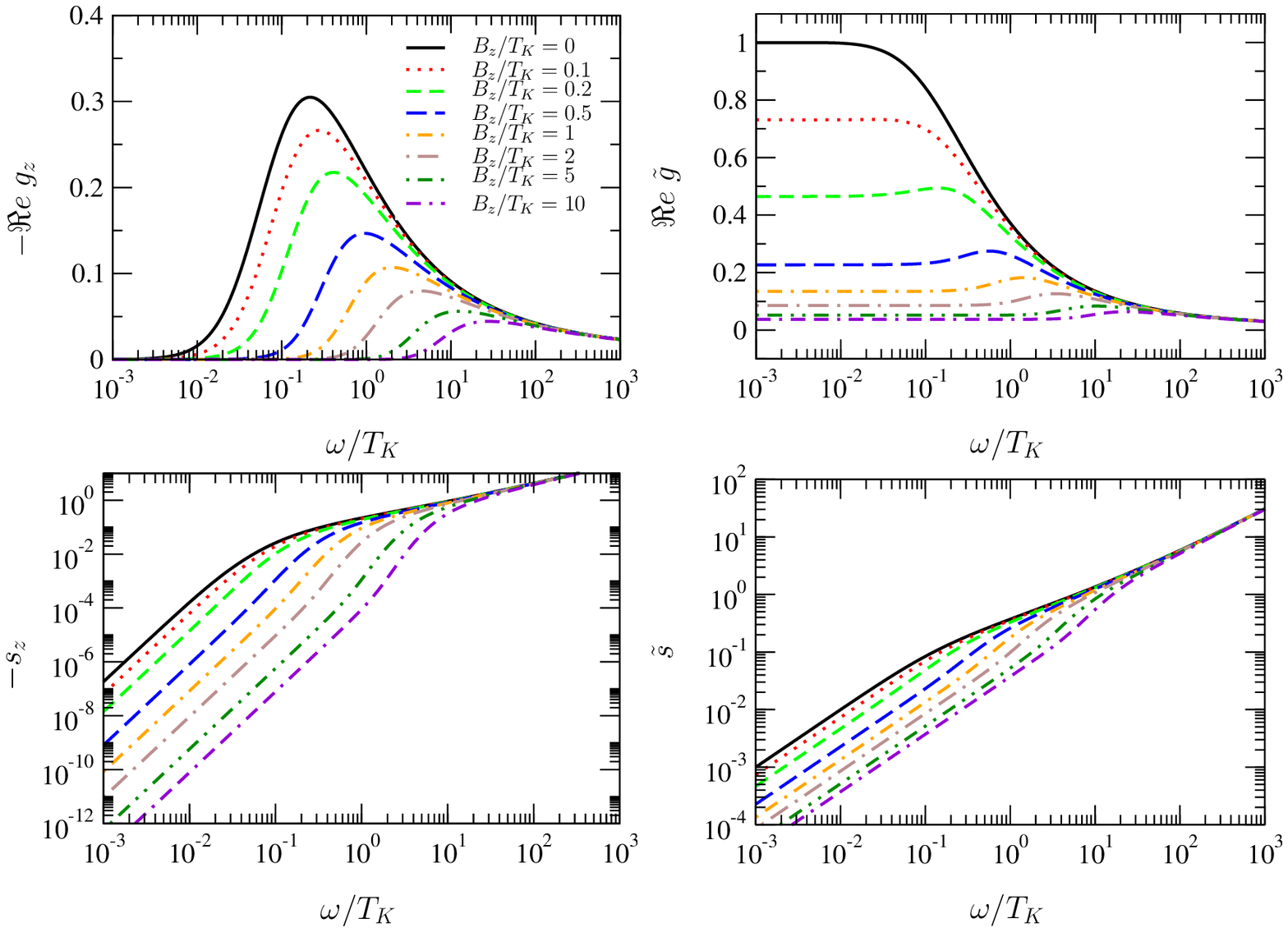}
    \end{center}
\caption{\label{fig:Bz}
(Color online)
Real parts of the universal functions 
$g_z$ and $\tilde g$, and the corresponding noise scaling functions,
$s_z$ and $\tilde s $, as obtained
by performing  zero-temperature NRG calculations with a
magnetic field $B_z$ along the $z$th direction.}
\end{figure*}

The behavior of the universal function, $-\myRe g
\sim G_{LR}^{\uparrow\downarrow}$, 
is somewhat more surprising. It exhibits a {\em maximum} at a frequency, 
$\w\sim T_K$, 
in agreement with the results of 
our analytical considerations, Eqs.~\eqref{g:largeomega}
and \eqref{g:smallomega}.
This maximum corresponds to a temporary spin
accumulation on the quantum dot at the ``resonance'' frequency,
$\omega \sim T_K$.

The lower panel of Fig.~\ref{fig:universal_functions} 
shows the universal functions for the spin noise 
on a logarithmic scale. The function $\tilde s$ shows a clear 
linear behavior at frequencies $\w<T_K$. The deviations at 
larger frequencies from the linear behavior 
are due to logarithmic corrections. 
At high frequencies $-s$ follows the behavior of 
$\tilde s$, and is linear apart from the aforementioned 
logarithmic corrections. 
For  $\w<T_K$, however, it deviates strongly, and scales 
to zero as $\sim \w^3$.

\subsubsection{Effect of Zeeman field, $B\ne 0$}

The presence of a Zeeman field, $H_B=-\bm{B}\cdot \bm{S}$, 
breaks the spin SU(2) symmetry. 
Therefore, for general spin polarizations, $\bm{n}$ and $\bm{n'}$,
the spin current noise and the spin conductance cannot be characterized 
by just two universal functions, as before.
Using group theoretical arguments and exploiting the 
electron-hole symmetry of the Kondo model
 we can show  that 5 universal functions need be used to characterize  
the complete $\bm{n}$ and $\bm{n'}$ dependence of $G_{rr'}^{\bm{n}\bm{n'}}$.
Moreover, these functions will not just be functions of 
$\w/T_K$ and  $T/T_K$, but they also depend on the ratio, $B/T_K$. 

Here we do not venture to characterize all these functions. Rather, we focus 
our attention to the special case, where  $\bm{n} \;|| \;\bm{n'} $.  
Without  loss of generality, we can then assume 
that $\bm{n}$ and $\bm{n'}$ are both parallel 
to the $z$ axis,  $\bm{n} =\s \;\hat z$ and $\bm{n'}  =\s' \;\hat z$. 
In this special case, one can show that, 
as a consequence of electron-hole symmetry and 
rotational invariance around the axis of the magnetic field, 
the contribution of the 
current $\tilde I^\s_r$ does not depend on the direction of the 
magnetic field, and that the total conductance 
has the structure, 
\bea 
&&G_{rr'}^{\s\s'}(\omega,\theta)
= 
{ {e^2\over h}  }
\;  
\Big \{ \delta_{\sigma\sigma'} \; {\tilde a}_{rr'} \; \tilde g  (\omega,B)
\\
&&+\;
  {\sigma\sigma'}\; {a}_{rr'} \;
\big[ \cos^2(\theta)\;g_z(\omega,B)  + \sin^2(\theta)\;g_\perp
(\omega,B)\big]
\Big\} \;,
\nonumber 
\eea
where $\theta$ is the angle between the magnetic field 
and $\bm{n}$, and two new scaling functions replace 
the original scaling function, $g(\omega,T)$. Notice that all these 
scaling functions now depend on the size of the magnetic field, too. 
Application of the fluctuation dissipation theorem yields then 
a similar scaling form for the noise, with corresponding scaling 
functions, $\tilde s$, $s_z$, and $s_\perp$, 
\bea 
&&S_{rr'}^{\s\s'}(\omega,\theta)
= -{e^2\over h}\;T_K  
\Big\{ \delta_{\sigma\sigma'} \; {\tilde a}_{rr'} \; \tilde s  (\omega,B) 
\\
&&+\;
  {\sigma\sigma'}\; {a}_{rr'} \;
\big[ \cos^2(\theta)\;s_z(\omega,B)  + \sin^2(\theta)\;s_\perp
(\omega,B)\big]
\Big \}\;.
\nonumber 
\eea

To determine the scaling functions, $\tilde g$ and $g_z$, and the 
corresponding noise scaling functions, we performed the zero-temperature NRG calculations 
for the case, where the magnetic field 
is parallel to the $z$ axis, $B_z = B$ (i.e., $\theta = 0$). The results are summarized in 
Fig.~\ref{fig:Bz}.
 Application of a magnetic field gradually removes the spin degeneracy 
of the dot, and leads to a splitting and suppression 
of the Kondo resonance for $B\gg T_K$. 
The splitting of the resonance can readily be 
 observed in the scaling function $\tilde g$ 
(i.e., the charge conductance through the dot). At very large fields, 
$B\gg T_K$, the conductance has a logarithmic tail 
for $\w\gg B$, but is 
strongly reduced at frequencies $\omega < B$, where spin-flip processes 
are forbidden.

Similar to $g$, the scaling function $g_z$ continues to vanish at $\omega=0$ for any 
magnetic field. This property follows  from the overall conservation 
of the total spin component parallel to the external magnetic field. 
The only effect of a magnetic field is thus to increase the 
 region of suppressed conductance.  

\begin{figure}
  \begin{center}
    \begin{minipage}[t]{0.9\columnwidth}
      \includegraphics[width=0.98\columnwidth, clip]{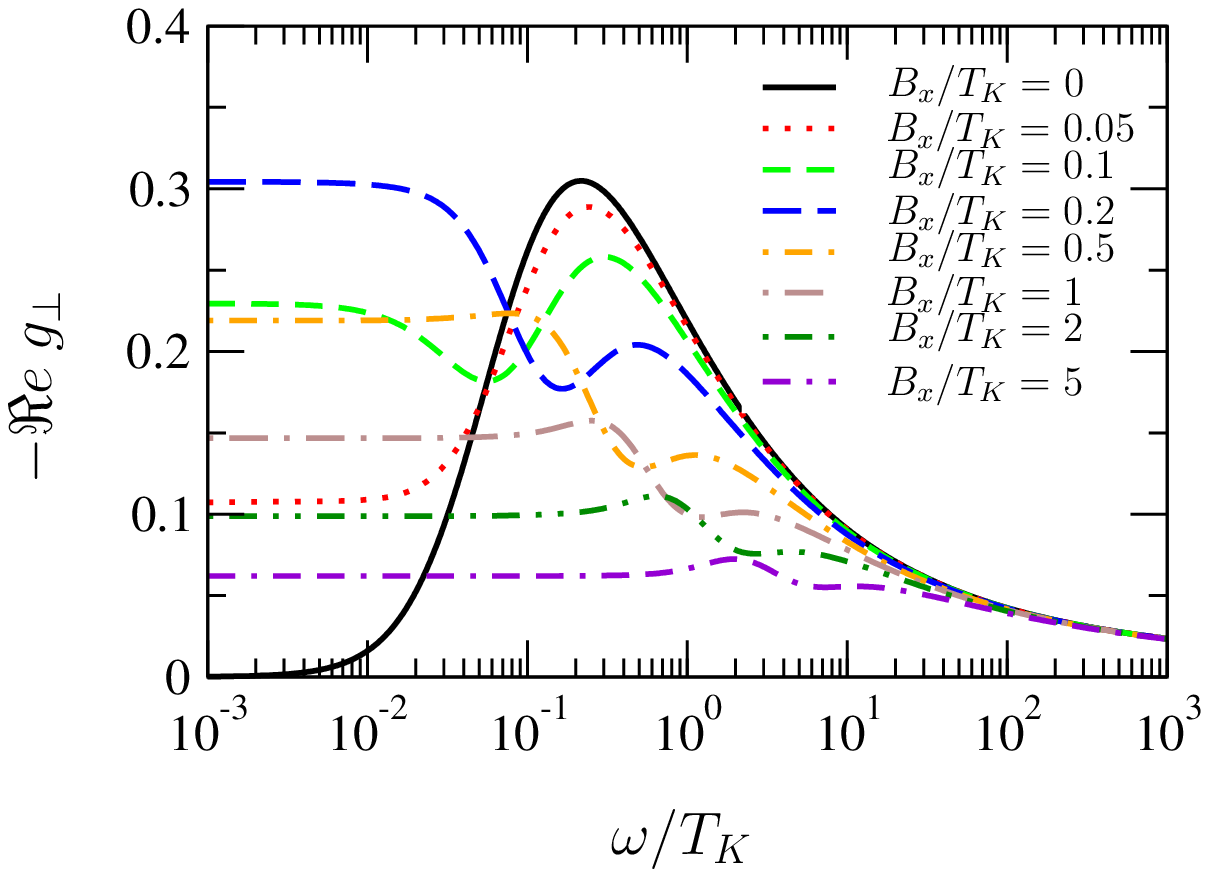}
    \end{minipage}\hfill
    \begin{minipage}[t]{0.9\columnwidth}
      \includegraphics[width= 0.98\columnwidth, clip]{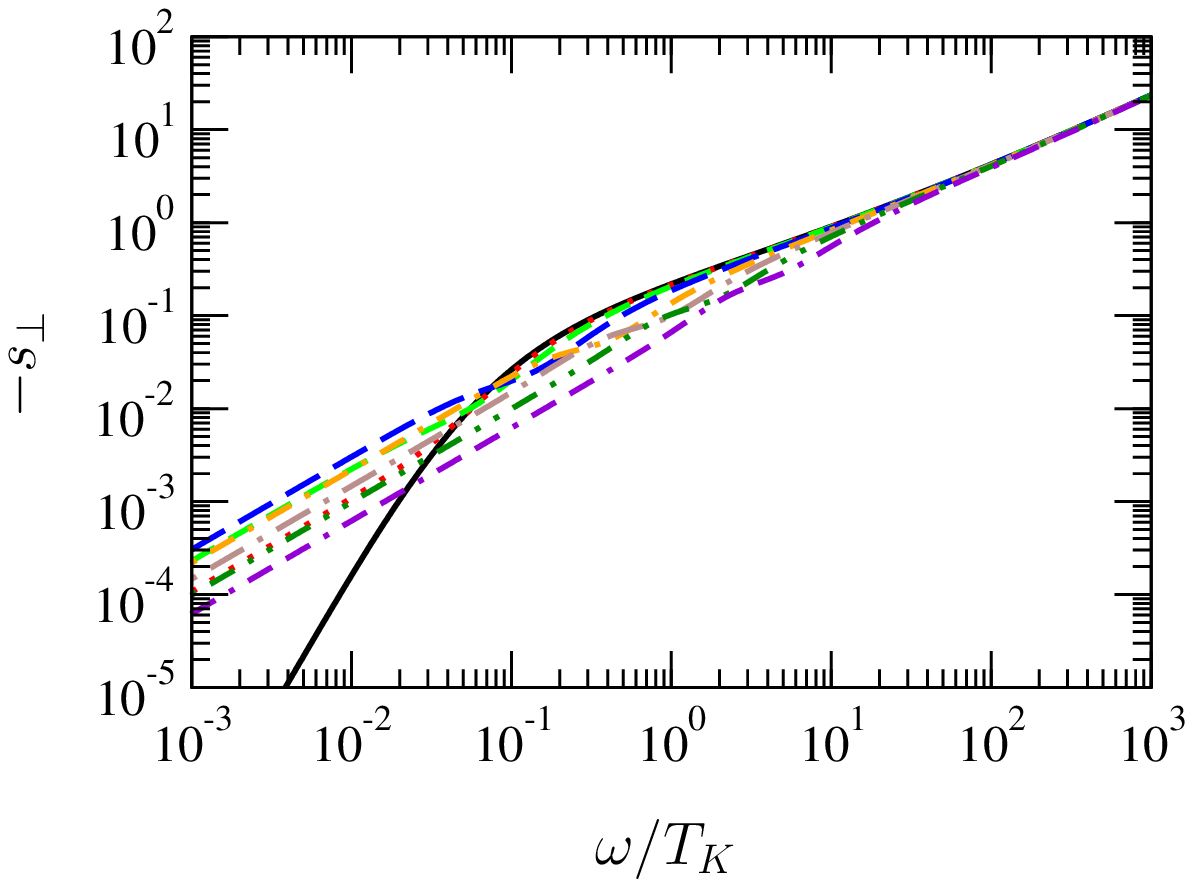}
    \end{minipage}
    \end{center}
\caption{\label{fig:Bx}
    (Color online)  Zero temperature NRG results for the real part
    of the universal function $g_\perp$, 
and the corresponding noise scaling function $s_\perp $
for several values of magnetic field pointing along the $x$ direction.}
\end{figure}

Next, to determine the scaling function, $g_\perp$, we performed calculations 
with a field $B_x$ applied along the $x$ direction ($\theta = \pi/2$). We also checked that 
$\tilde g$ is the same as before, and does not depend on the direction
 of the applied field. The function  $g_\perp$, however,  is markedly different from 
$g_z$. Of course, for $B=0$ they are both equal to $g$.
However, any finite magnetic field results in a finite d.c. conductance, $g_\perp(\omega=0)\ne 0$. 
The reason of this is that in this case we measure a spin component which is 
not conserved for any finite magnetic field. 
In a  sense, the perpendicular magnetic field in 
this case plays a role similar to the external spin relaxation, discussed 
in Section~\ref{sec:spin_relaxation}, and allows the impurity spin to flip 
back and forth, independently of the conduction electrons. 
As a consequence, the "pseudogap" feature 
in $g_\perp$ is gradually filled up with increasing $B_x$,
reaching maximum for $B_x \sim T_K$.


\section{The finite temperature case}

Although the NRG calculations presented in the previous section could, 
in principle be extended to any finite temperature, 
a serious {\em technical} problem appears. As we mentioned in the 
introduction, the cross-spin conductance, $G_{LR}^{\uparrow\downarrow}$
and thus $g$, must vanish at any finite temperature $T$ in the 
$\w \to 0$ limit. However, currently used finite temperature 
NRG broadening schemes all produce linear ($\propto \w$) spectral weight 
in $\varrho_{{\cal I}_\uparrow {\cal I}_{\uparrow}}(\omega, T)$, and therefore 
by Eq.~\eqref{eq:Reg_NRG_expression} lead to a finite 
and thus unphysical d.c. cross-spin conductance. 

We do not know of any way to get around this problem. 
Therefore, at finite $T$, 
we had to rely on analytical results, and combine them 
with the  $T=0$ temperature results to obtain a coherent picture. 
At very high temperatures,  $T\gg T_K$, we can make use of
perturbative approaches. However, as explained below, 
even in this regime simple-minded perturbation theory is 
insufficient, and we need to combine it with 
a master equation  approach to obtain the complete $\w$ dependence of the 
spin conductance and noise. Combining these perturbative results 
with scaling and Fermi liquid arguments, we are then able to 
understand the complete frequency and temperature dependence 
of the scaling functions, $s$, $\tilde s$, $g$ and $\tilde g$.

\subsection{Perturbation theory and master equation approach}

At temperatures $T\gg T_K$ corrections to the leading order 
perturbative results are small, and
 much of the noise spectrum can be understood 
based upon perturbative results. However, 
to understand the limitations of perturbation theory,
 we first need to understand the 
important time scales in this high temperature limit
and the way they influence spin transport.
At $T\gg T_K$, transport through the dot occurs 
through individual exchange processes, whereby just one electron 
tunnels from one side of the dot to the other side of it. 
The typical time between such events is given by the "Korringa time", 
$\tau_K$, which we define as the inverse of the
Korringa rate, $\tau_K\sim h/E_K$.
To the leading order in $j$, it is given as, $ \tau_K\sim h /j^2 T$. 
Tunneling events are, however, not instantaneous in the sense 
that they are dressed by the internal dynamics of the 
electron-hole excitations, created throughout the tunneling process. 
Correspondingly, the "duration" of a tunneling process is given 
by the thermal time, 
{$\tau_T\equiv h/T\ll \tau_K$}. 
At very short times below the thermal time, 
$t< h/T\equiv \tau_T$ (or at frequencies { $\w> T$}), 
current-current correlations reflect just the internal and coherent dynamics 
of such a spin-flip event, as well captured by the usual 
connected  second order contribution to the current-current
correlation function. 

As just stated, the usual bubble 
diagram  only accounts for  the structure of a {\em single} tunneling event. 
At times $t>\tau_K$, however, several independent incoherent 
tunneling processes take place. These processes are correlated, since 
a spin-flip process that changes the dot spin from up into down
($\Uparrow\rightarrow \Downarrow$)
must necessarily be followed
by an opposite process when the dot spin flips from down to up ($\Downarrow\rightarrow \Uparrow$).
These correlations turn out to be important for spin transport, and
are obviously not captured by simple perturbation theory. 
Fortunately, for times $t>\tau_T$ ($ \w< T$), 
the internal dynamics of a spin-flip event can be ignored, 
and we can make use of  a master equation method as a  
complementary approach. There tunneling processes are taken to 
be instantaneous, just characterized by some rates, but  correlations 
between individual spin-flip events are properly accounted for 
through a classical rate equation. 

From these simple arguments we thus conclude that the master 
equation approach must be valid for frequencies $\w< T$, while 
simple-minded perturbation theory works  for frequencies
$ E_K< \w$. Since $E_K< T$, the range of validity 
of these two approaches overlaps, as also confirmed by the 
actual calculations presented below and by the results of 
Ref.~[\onlinecite{KindermannPRB2005}].

\subsubsection{Perturbation theory}
\label{sec:perturbation_theory}

As explained before, for $T\gg T_K$ and 
times $t< \tau_K$ (frequencies $\w> E_K$), 
perturbation theory (PT) is thus a good approximation, 
though it fails at longer times where already several spin-flip events 
occur, and  the correlations between these spin-flip events cannot 
be neglected. 
Simplest 0-th order perturbation theory 
yields, e.g., for the left-right components 
of the symmetrized frequency-dependent spin noise 
\bea
  S_{LR}^{\uparrow \downarrow}(\w) &=& - \frac{e^2}{h} \;
\sin^2\phi \;  \frac{\pi^2 j^2 }{8}\;
\omega\;\coth\left(\frac{\omega}{2T}\right) + \dots\;,
\nonumber
\\
  S_{LR}^{\uparrow \uparrow}(\w) &=& - \frac{e^2}{h } \;
\sin^2\phi \;  \frac{\pi^2 j^2 }{16} \;
\omega\;\coth\left(\frac{\omega}{2T}\right) + \dots\;,
  \nonumber
\eea 
with the dots referring to higher order corrections in $j$. 
The corresponding universal functions then read 
for $\w > E_K$,
\bea
 {s}^{\;\rm PT}(\w) &=& - \frac 2 3 \; \tilde   s^{\;\rm PT}(\w) = - \frac{\pi^2 j^2 }{8} \;
\frac \omega {T_K}\;\coth\left(\frac{\omega}{2T}\right) + \dots\;, 
\nonumber
\\
 g^{\;\rm PT}(\w) &=& - \frac 2 3 \; \tilde   g^{\;\rm PT}(\w) = - \frac{\pi^2 j^2 }{8} \;
+ \dots\;. 
\label{eq:s_perturbative}
\eea 
Higher order terms give  logarithmic corrections, and 
lead to a renormalization of $j$ in these expressions.

\subsubsection{Master equation approach}
\label{sec:master_equation}

Let us now focus on the "classical" 
frequency regime, $\w<T$. Here we can use 
a simple master equation (ME) approach,~\cite{korotkov} where 
we assume  that, at any instance, the  spin
on the dot is either in a spin-up state $S = \;\Uparrow$  or in 
a spin-down state  $S = \;\Downarrow$, with corresponding probabilities, ${\cal P}_S(t)$.
Conduction through the dot and spin relaxation 
are generated by scattering events, $q$, generated by the exchange interaction, 
Eq.~\eqref{eq:H_int}.
These scattering events are taken to be  instantaneous, and consist 
of the scattering of  a spin $\s$ electron from lead $r$ 
to a spin $\s'$ state in lead $r'$
while changing the dot spin, $S\to S'$, 
$$
q\leftrightarrow \{r',\s',S' \leftarrow r,\s,S\}.
$$ 
They  
occur with a rate, $\gamma(q) = \gamma_{r'\s'\leftarrow r\s}^{S'\leftarrow S}
\propto j^2 v_r^2 v_{r'}^2$, following Fermi's golden rule. 

The dynamics of the dot spin is described by a simple
master equation, 
\begin{equation}
\frac{\rm d}{{\rm d}t}
\left ( \begin{array}{c}
 {\cal P}_\Uparrow \\  {\cal P}_\Downarrow
 \end{array} \right)
 = 
\left ( \begin{array}{cc}
 -\Gamma   & \Gamma \\
 \Gamma    & - \Gamma 
  \end{array} \right) 
\left (  \begin{array}{c}
 {\cal P}_\Uparrow \\
 {\cal P}_\Downarrow
 \end{array}  \right)
\,,
\label{eq:spinmaster}
\end{equation}
with the relaxation rate $\Gamma$ given as
$\Gamma  \equiv \sum_{r,r',\s,\s'} \gamma_{r\s\leftarrow
  r'\s'}^{\Uparrow  \leftarrow \Downarrow}$. 
From Eq.~\eqref{eq:spinmaster} it follows that spin polarization on
the dot relaxes exponentially, $\langle S_z\rangle\sim e^{-2\Gamma t}$.  Thus the rate $\Gamma$ is related to
the Korringa spin relaxation rate as,  
$E_{K,0} \equiv 2 \;\Gamma$,  which  for the simple equilibrium case 
considered here takes on  the usual expression, 
\be
E_{K,0} = 2 \;\Gamma = \pi \;j^2 \; T\;. 
\label{eq:Korringa_0}
\ee
Here the label "$0$" indicates that this 
is just the leading order expression of the Korringa rate, 
and higher order terms in perturbation theory 
renormalize it [see Eq.~\eqref{eq:Korringa}].

To compute  current-current correlations, we first 
notice that a scattering event $q$ at some time $\tau$ induces current pulses 
in the leads, 
$J_{r}^{\s}(t) = {\Delta Q}_{r}^{\s}(q)\;\delta(t-\tau)$,
with $ {\Delta Q}_{r}^{\s}(q)\in \{\pm e,0\}$ 
the amount of charge transferred. 
Similarly,  a series of events, $\{\tau_n, q_n \}$ generates a current,
\be
J_{r}^{\s}(t) = \sum_n {\Delta Q}_{r}^{\s}(q_n)\;\delta(t-\tau_n)\;. 
\ee
As a consequence, the stationary current-current correlation function can 
be simply expressed as 
\begin{widetext}
\begin{equation}
  \langle J_{r}^{\s}(t)\, J_{r'}^{\s'}(0)   \rangle
  = \delta(t) \sum_q P(q) \; {\Delta Q}_{r}^{\s}(q) {\Delta Q}_{r'}^{\s'}(q)+
  \sum_{q,q'}P(q,t\;; \;q',0)\;{\Delta Q}_{r}^{\s}(q)\; {\Delta
    Q}_{r'}^{\s'}(q') \,.
\label{eq:semi_current_current}
\end{equation}
\end{widetext}
Here the first term describes the auto-correlation of individual scattering 
events, while the second term describes correlations between distinct tunneling 
events. The probability $P(q)$ in Eq.~\eqref{eq:semi_current_current}
denotes the stationary rate for 
a given type of event, { $q =\{ r',\s',S'\leftarrow r,\s,S  \}$}, 
and can be expressed as
\be 
P(q) = \gamma(q)
 \;\overline {\cal P}_{S}\;,
\ee
with $ \overline {\cal P}_{S}$ the stationary probability 
distribution of the dot spin. 
The quantity   $P(q,t\; ;\; q',0)$ denotes the joint 
probability rate of a scattering  event  $q$ at time $t$ and an event $q'$ 
at time $t=0$. For the events
{ $q=\{ r_2,\s_2,S_2\leftarrow r_1,\s_1,S_1\} $}
and { $q'=\{ r_2',\s_2',S_2' \leftarrow r_1',\s_1',S_1'  \} $}
it can be expressed  as
\begin{equation}
  P(q,t\; ;\; q',0) = \gamma(q)\; P_{S_1\leftarrow S_2'}(t)\;
\gamma(q') \;\overline {\cal P}_{S_1'}\,, 
\end{equation}
where $ {\cal P}_{S_1\leftarrow S_2'}(t)$ denotes the conditional 
probability that the dot spin evolves from state 
$S_2'$ to $S_1$ during time $t$. The function  
$ {\cal P}_{S_1\leftarrow S_2'}(t)$ is determined 
by the master equation, Eq.~\eqref{eq:spinmaster}, 
and it is obviously this and only this 
quantity that generates 
time-dependent correlations between consecutive scattering events. 
Thus spin current correlations in this perturbative
master equation approach are directly related 
to the time evolution of the dot spin. 

\begin{figure}[t]
  \includegraphics[width=0.95\columnwidth,clip]{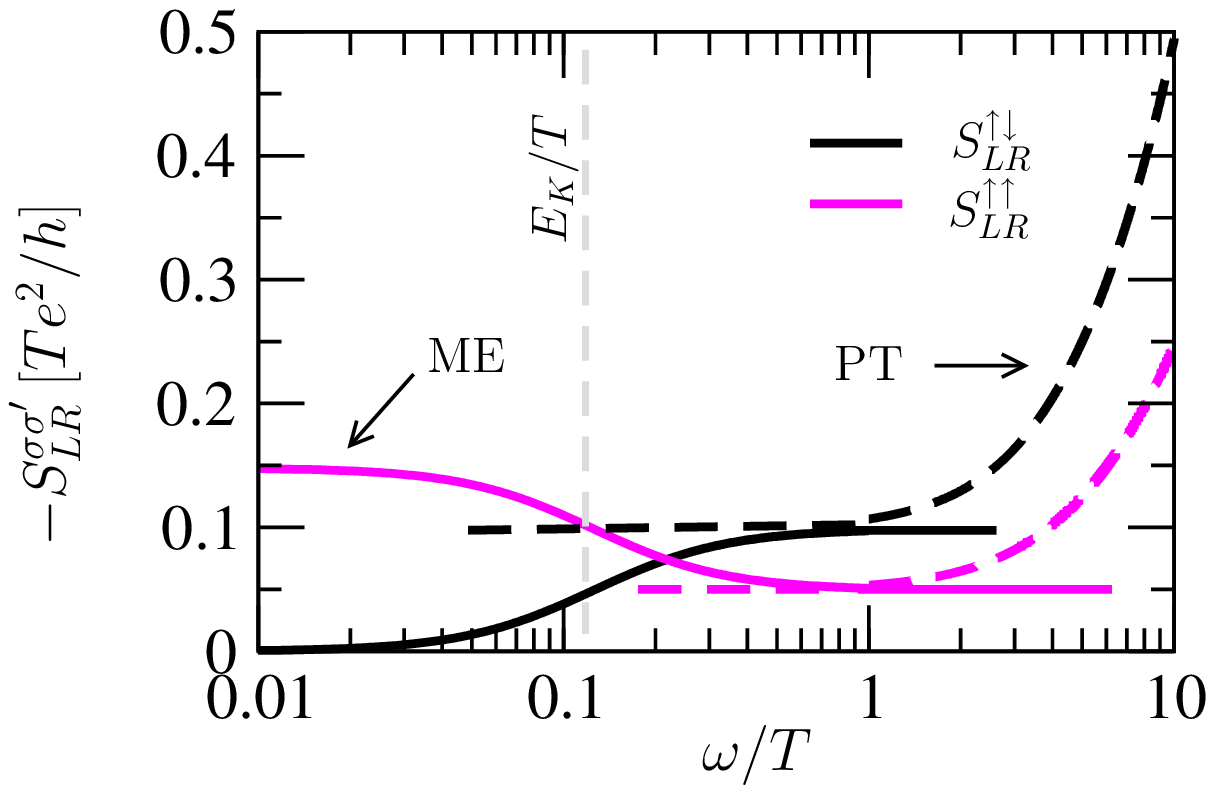}
  \includegraphics[width=0.95\columnwidth,clip]{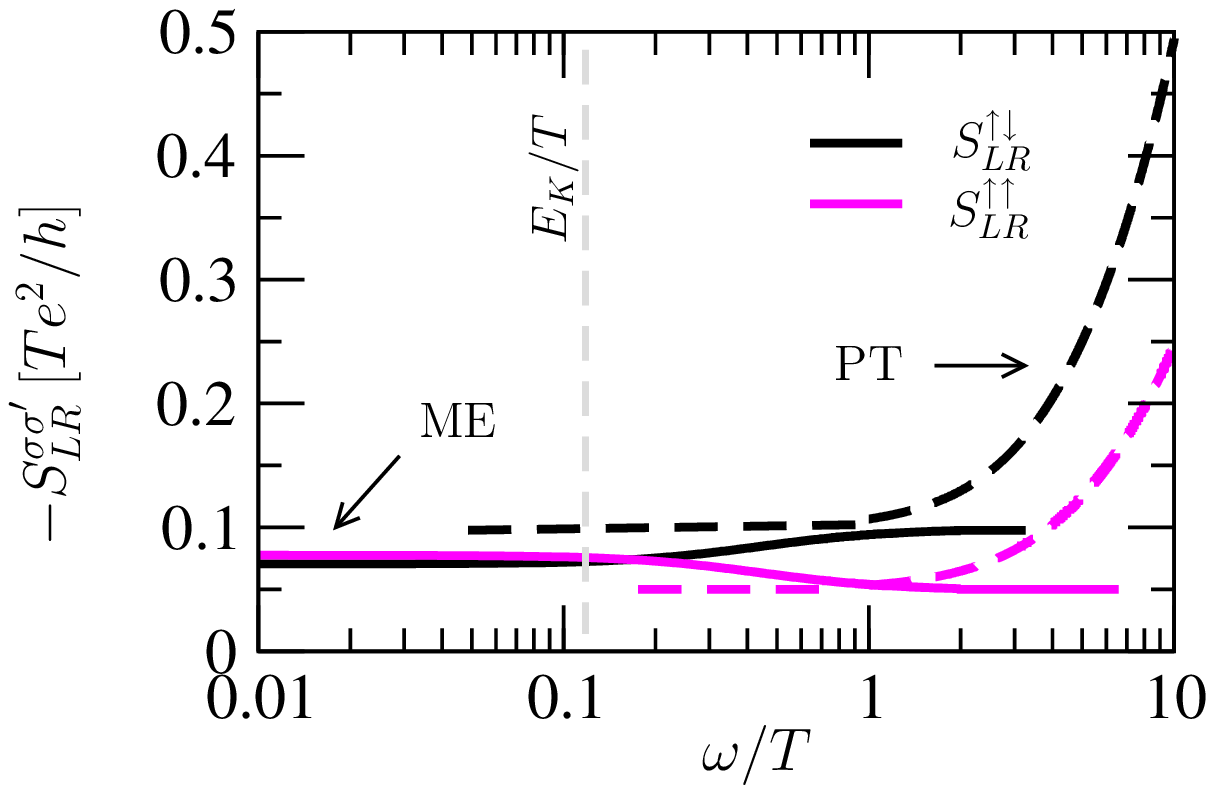}
  \caption{\label{fig:shot_noise_finite_T_comparison}
  (Color online) Upper panel: Frequency-dependent
  spin current cross-correlations ($S_{LR}^{\uparrow\downarrow}$) 
  and auto-correlations ($S_{LR}^{\uparrow\uparrow}$) in the absence of 
  spin relaxation, as calculated 
  using the master equation approach (ME, solid line) and
  second-order perturbation theory (PT, dashed line).
  In the calculations we assumed $j=0.2$. 
  Lower panel: Effect of a finite (rather large)
  external spin relaxation rate, $1/\tau_s \equiv E_{K,0}/2 $.
}
\end{figure}

Having set up this general framework, 
the detailed calculation of the classical noise spectrum 
is somewhat tedious, but straightforward. Therefore, 
instead of presenting further details on it, let us just 
continue with the discussion of  the final results. 
Within the master equation approach, the 
left-right components of the spin noise read 
\bea 
S_{LR}^{\uparrow \downarrow} (\omega< T) &\approx&
 -\frac{e^2}{h }\;\frac{\pi}{4}  \frac{E_{K, 0}\; \omega^2
\sin^2 \phi} {\omega^2+E_{K,0}^2}\;,
\label{eq:noise_master}
\\
S_{LR}^{\uparrow \uparrow} (\omega<T)&\approx & 
  -\frac{e^2}{h}\; \frac{\pi}{8}
  \frac{E_{K,0}\, \left (\omega^2 + 3\; E_{K,0}^2 \right)\sin^2 \phi}
  {\omega^2+ E_{K,0}^2}\;.
\nonumber
\eea 
From these equations we extract  the following 
approximations for the universal scaling functions, 
\bea
s^{\;\rm ME}(\w) &=& - \frac{\pi^2}{4} \;j^2 \frac{T}{T_K}\;
\frac{\omega^2} {\omega^2+E_{K,0}^{\;2}}\;,
\\
\tilde s^{\;\rm ME}(\w)& = &   \frac{3 \pi^2}{8}\;j^2\; \frac{T}{T_K} 
\;.
\eea
Remarkably, for $T\gg \w \gg E_{K,0}$, these 
results precisely coincide with the 
perturbative results, Eq.~\eqref{eq:s_perturbative}. 
This is indeed also clearly visible in 
Fig.~\ref{fig:shot_noise_finite_T_comparison}, where 
we compare perturbation theory results for $S_{LR}^{\s\s'}(\w)$
with results of the  master equation approach. 
We remark that one can bridge these two approaches through 
a systematic but 
much more formal and difficult quantum Langevin 
approach, already briefly sketched in Ref.~[\onlinecite{Rapid}], 
and to be discussed in a subsequent publication, Ref.~[\onlinecite{future}].

The fluctuation dissipation theorem, 
Eq.~\eqref{eq:FDT}, in this classical 
regime, $\w< T$, amounts in  the following  
expressions, 
\bea
\myRe g^{\;\rm ME}(\w) &\approx&\frac{T_K}{2T} \;s^{\;\rm ME}(\w) =  - \frac{\pi^2}{8} \;j^2 \;
\frac{\omega^2} {\omega^2+E_{K,0}^2}\;,
\\
\myRe \tilde g^{\;\rm ME}(\w)& \approx & \frac{T_K}{2T}\; \tilde s^{\;\rm ME}(\w) = 
 \frac{3 \pi^2}{16}\;j^2\;.
\eea
Notice that, in contrast to  $s$ and $g$,
the functions  $\tilde s$ and $\tilde g$
are completely featureless in this frequency range. 
On the other hand, in agreement with our earlier statement,  $S_{LR}^{\uparrow
\downarrow}(\omega)$, $\myRe G_{LR}^{\uparrow
\downarrow}(\omega)$, $s$ and $\myRe g$ all 
exhibit a dip below $E_K$, and scale to zero as $\omega\to0$. 
There is a simple heuristic 
picture behind this fact: a spin-flip 
process $\Uparrow\to \Downarrow$ pumps spin-up electrons into
the leads. However, it must necessarily be followed by a reverse 
spin-flip process, $\Downarrow \to \Uparrow$, where, on average,
exactly the same amount of spin is pumped back as it has been pumped in 
before. These processes exactly balance each other 
in the long time limit, and lead to a vanishing cross-spin 
conductance in equilibrium. We remark that if, however, there are external spin 
relaxation processes, then the spin may flip back spontaneously 
before pumping back the injected spin through the reverse 
spin-flip process. In this case, as we shall see in 
Sec.~\ref{sec:spin_relaxation}, the conductance  $G_{LR}^{\uparrow
\downarrow}(\omega)$ remains finite even in the $\w\to 0$ limit
(see also Fig.~\ref{fig:shot_noise_finite_T_comparison}).  

A rather curious consequence of the dip in 
$s$ is that, while  $S_{LR}^{\uparrow \downarrow}(\omega)$ 
develops a  {\em dip} below the Korringa rate, the noise component
$S_{LR}^{\uparrow \uparrow}(\omega)$ exhibits a {\em peak} of
equal size, which precisely cancels the dip 
of  $-S_{LR}^{\uparrow \downarrow}(\omega)$ 
in the charge noise.
This peak in $S_{LR}^{\uparrow \uparrow}(\omega)$
or the similar peak in  $S_{LL}^{\uparrow \uparrow}(\omega)$
 may be more conveniently detected experimentally than 
 cross-spin correlations.

\subsection{Beyond perturbation theory}

\begin{figure}[t]
  \includegraphics[width=0.95\columnwidth]{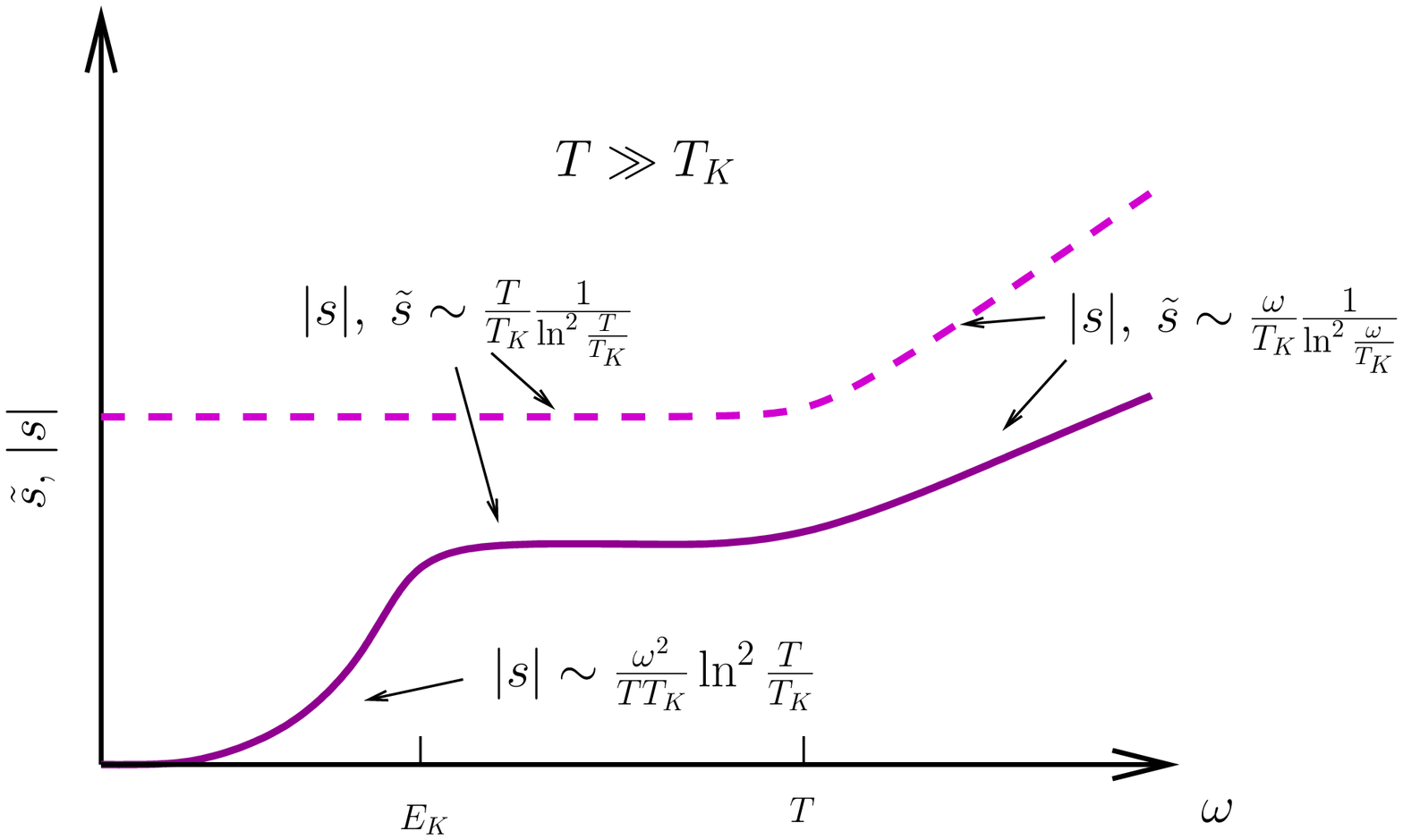}
  \includegraphics[width=0.95\columnwidth]{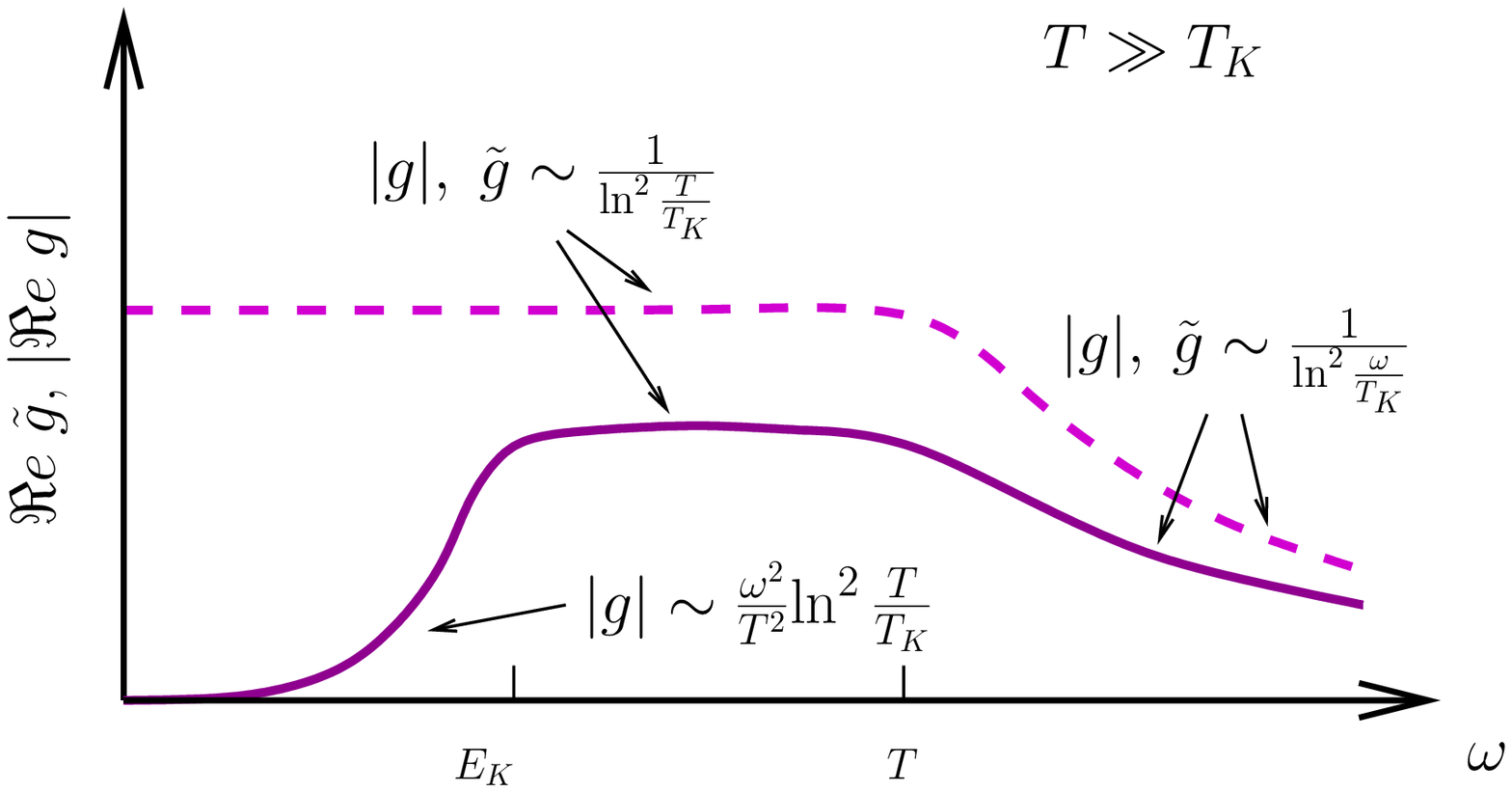}
  \caption{\label{fig:omega_T_gg_TK}
  (Color online)   Sketch of the universal scaling functions $s$ (continuous line),
  and $\tilde s $ (dashed line), and the real parts of 
 $g$ (continuous line) and $\tilde g(\omega/T_K, T/T_K)$ (dashed line)
for $T\gg  T_K$. The charge conductance and noise functions, 
$\tilde g$ and $\tilde s$ show no particular feature, while the spin 
scaling functions exhibit an anomaly below the Korringa rate, $E_K$. }
\end{figure}

\subsubsection{Logarithmic corrections}

In the previous subsection we discussed only
the leading order perturbative and master 
equation results. Performing, however, perturbation theory 
in $j$ gives rise to logarithmic corrections. As long as 
$\max\{T,|\omega|\} \gg  T_K$, these 
corrections can be summed up using  
renormalization group methods,~\cite{Abrikosov,FowlerZawadowski} 
and amount in the replacement of $j$ by it's renormalized value, 
$j \to 1/ {\ln(\max\{T,|\omega|\} / T_K)}
\label{eq:renormalized_j}$.
Apart from this substitution, however, 
the results of 
Subsections~\ref{sec:perturbation_theory} and 
\ref{sec:master_equation} continue to be valid as long as 
$T\gg T_K$. For $\w>T$, e.g., we just recover the $T=0$ 
temperature results, Eqs.~\eqref{g:largeomega} and \eqref{s:largeomega},
while in the opposite limit, $\w<T$,  we obtain 
\bea
s(\w) &\approx& - \frac{\pi^2}{4} \;\frac{1}{\ln^2(T/ T_K) }
\;\frac{T}{T_K}\;
\frac{\omega^2} {\omega^2+E_K^{\;2}(T)}\;,
\nonumber
\\
\tilde s(\w)& \approx  &   \frac{3 \pi^2}{8}
\;\frac{1}{\ln^2(T/ T_K) }
\; \frac{T}{T_K} 
\;, 
\eea
with  $E_K=E_K(T)= \pi T /\ln^2(T/ T_K) $ the renormalized 
Korringa rate of Eq.~\eqref{eq:Korringa}. Similarly, for the 
scaling functions $g$ and $\tilde g$ we obtain in this regime, 
\bea
\myRe g(\w) &\approx&\frac{T_K}{2\;T} \;s(\w) =  - \frac{\pi^2}{8} 
\;\frac{1}{\ln^2(T/ T_K) } \;
\frac{\omega^2} {\omega^2+E_K^{\;2}(T)}\;,
\nonumber
\\
\myRe \tilde g(\w)& \approx & \frac{T_K}{2\;T}\; \tilde s(\w) = 
 \frac{3 \pi^2}{16}\;\frac{1}{\ln^2(T/ T_K) }\;.
\eea
Fig.~\ref{fig:omega_T_gg_TK} gives a concise summary of these results.

\subsubsection{Fermi liquid regime, $T\ll T_K$}

\begin{figure}[t]
  \includegraphics[width=0.95\columnwidth]{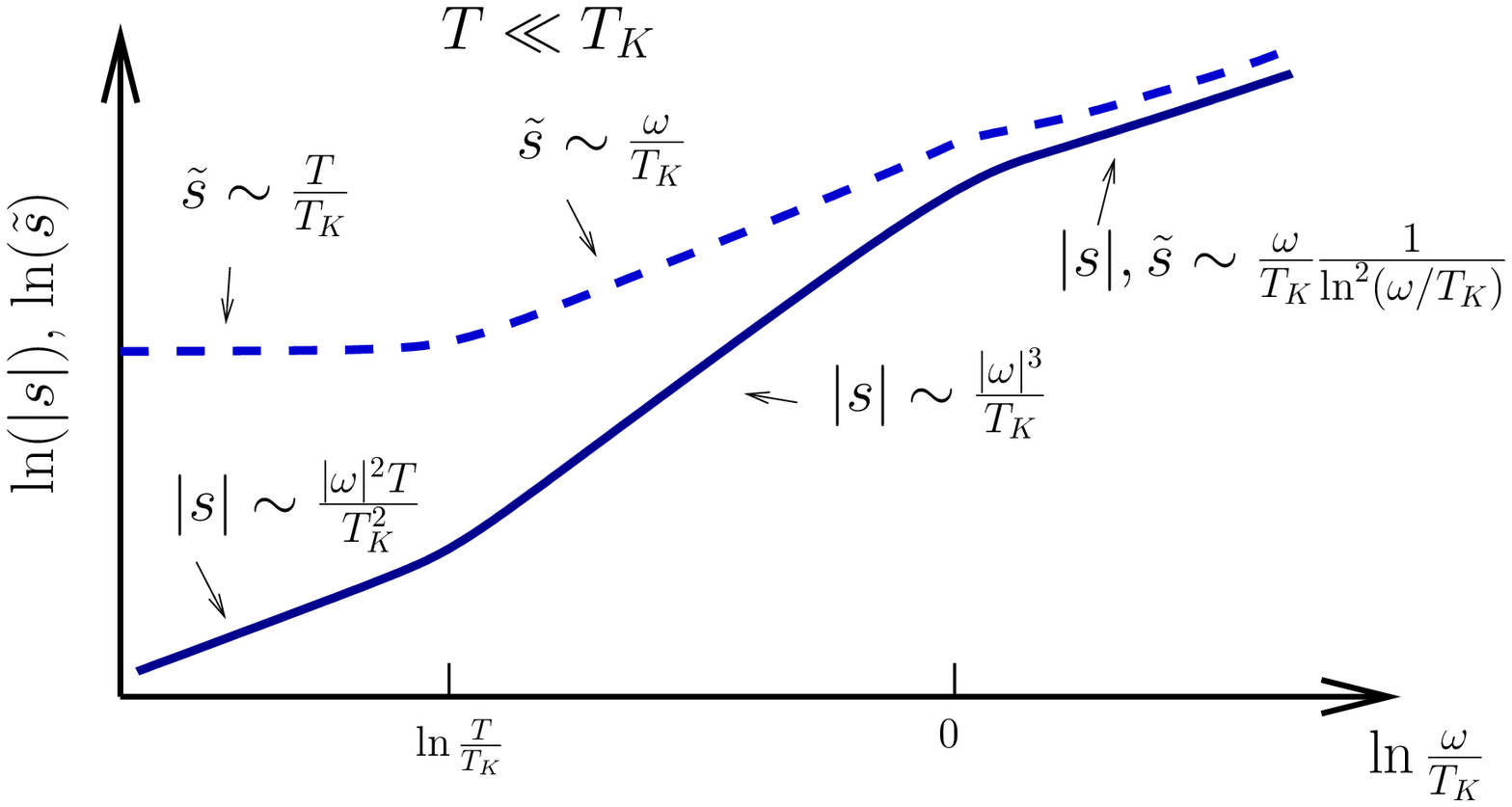}
  \includegraphics[width=0.95\columnwidth]{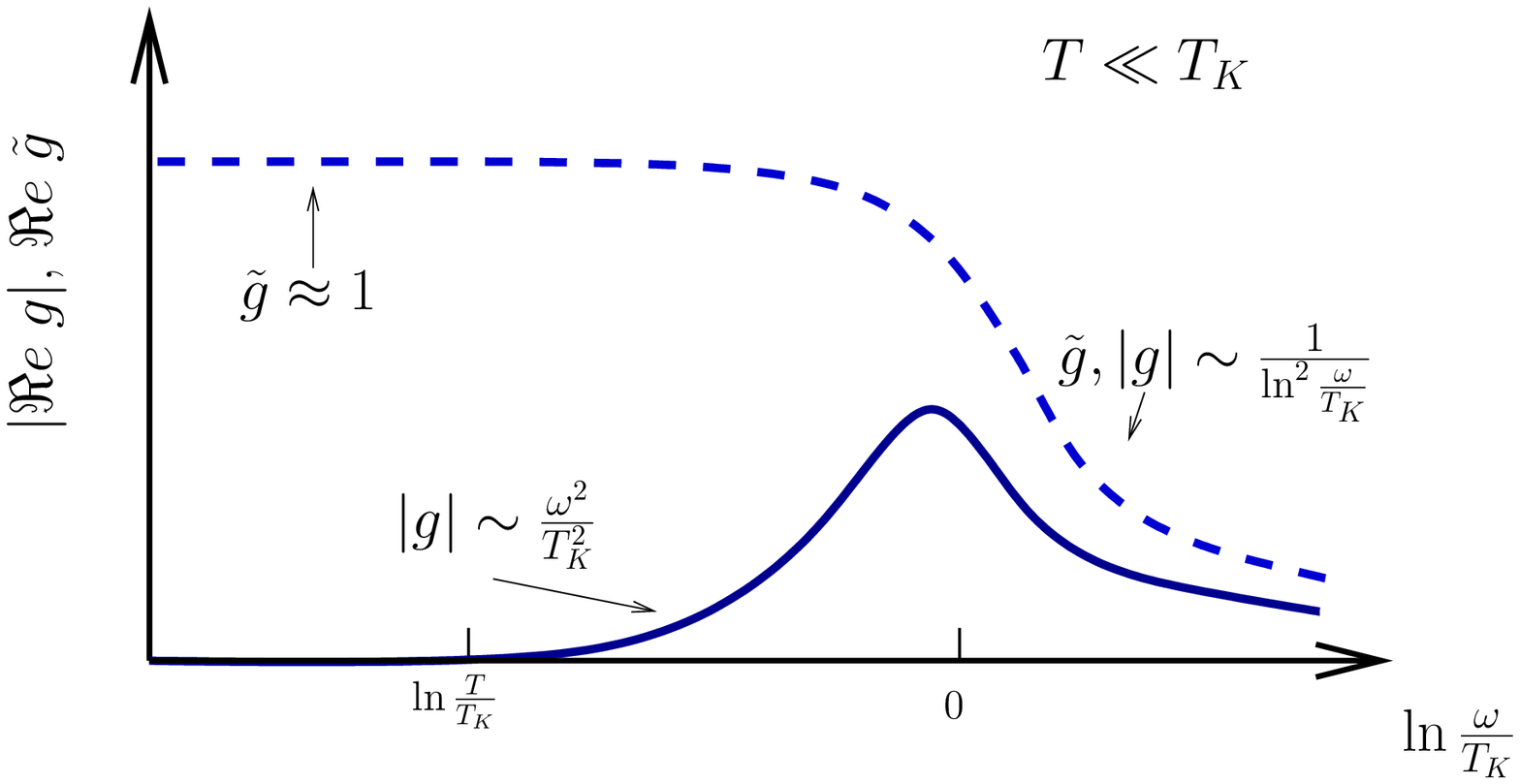}
  \caption{\label{fig:omega_T_ll_TK}
  (Color online)   Sketch of the universal scaling functions 
$s$ (continuous line) and  
$\tilde s$ (dashed line)  
and the real parts of  $g$ (continuous line) and 
$\tilde g$ (dashed line) for $T\ll T_K$. 
}
\end{figure}

In the Fermi liquid regime,~\cite{Nozieres} $T\ll T_K$, 
perturbation theory in $j$ breaks down. However, we can  
derive the behavior of the scaling functions by two simple 
observations.
We first observe that in this Fermi liquid regime, 
$j\to \infty$,~\cite{WilsonRMP75,Nozieres} and therefore 
the only remaining energy scales are $T$ and $T_K$. 
Our second observation is that at the Fermi liquid fixed point, 
the residual electron-electron interactions 
are irrelevant,~\cite{Nozieres}
and therefore physical quantities are analytical functions of 
$\omega$.  In particular, the asymptotic forms, 
Eqs.~\eqref{eq:tilde_g_FL} and \eqref{g:smallomega}
remain valid up to the higher order corrections even at finite temperatures,  
\bea 
\myRe g(\omega)&=& -\alpha  \frac{\w^2}{T_K^2} + {\cal O}(\w^4, T^2 \w^2)\;,
\\
\myRe \tilde g (\omega) &=& 1 +  {\cal O}(\w^2, T^2)\;.
\eea
The scaling functions of the noise, $s$ and $\tilde s$ can then yet again
be read out of the fluctuation dissipation theorem, Eq.~\eqref{eq:FDT}, 
yielding
\bea 
s(\omega\ll T_K)&\approx & -\alpha  \frac{\w^3}{T_K^3} 
\; {\rm coth}\left(\frac \omega {2T}\right)\;,
\\
\tilde s (\omega\ll T_K) &\approx& 
\frac \omega {T_K} \; {\rm coth}\left(\frac \omega {2T}\right)\;.
\eea
These equations reduce to the  $T=0$ expressions in the 
$\w\gg T$ limit, and  are valid as long as $\w<T_K$. 
Notice that in the $\w\to 0$ limit, the charge noise component 
$\tilde s$ scales to a constant, $\propto T$, while the 
spin component $s$ scales quadratically to zero, just as in 
the $T\gg T_K$ regime. Of course, for $\w> T_K$ the scaling functions 
must also reduce to their $T=0$ temperature expressions, 
Eqs.~\eqref{g:largeomega} and \eqref{s:largeomega}. The overall behavior 
of these functions for $T\ll T_K$ is 
sketched in Fig.~\ref{fig:omega_T_ll_TK}

\section{Transient response}
\label{sec:transient}

Let us now turn to the discussion of real time transient 
response, i.e., the time dependent current response in the left lead 
when a spin-dependent voltage
of the form  $V_{R}^{\uparrow}(t) = \delta V_{R}^{\uparrow}\; \theta (t)$ 
is applied to the right electrode (see Fig. \ref{fig:filters}). 
Within linear response theory, the average current pulse
is just given by 
\begin{equation}
  \expect{ J_{L}^{\s}(t)} =  i \int_{-\infty}^\infty G_{LR}^{\s\uparrow}
  (\omega)  \;
\frac1 {\w- i \delta}\;
{\rm e^{-i\,\omega\,t}}\; {\rm d}\omega \; \delta V_{R}^{\uparrow} \,,
\label{eq:current_time}
\end{equation}
with $i/(\w- i \delta)$ the Fourier transform of the $\theta$ function.

\begin{figure}[t]
  \includegraphics[width=0.95\columnwidth]{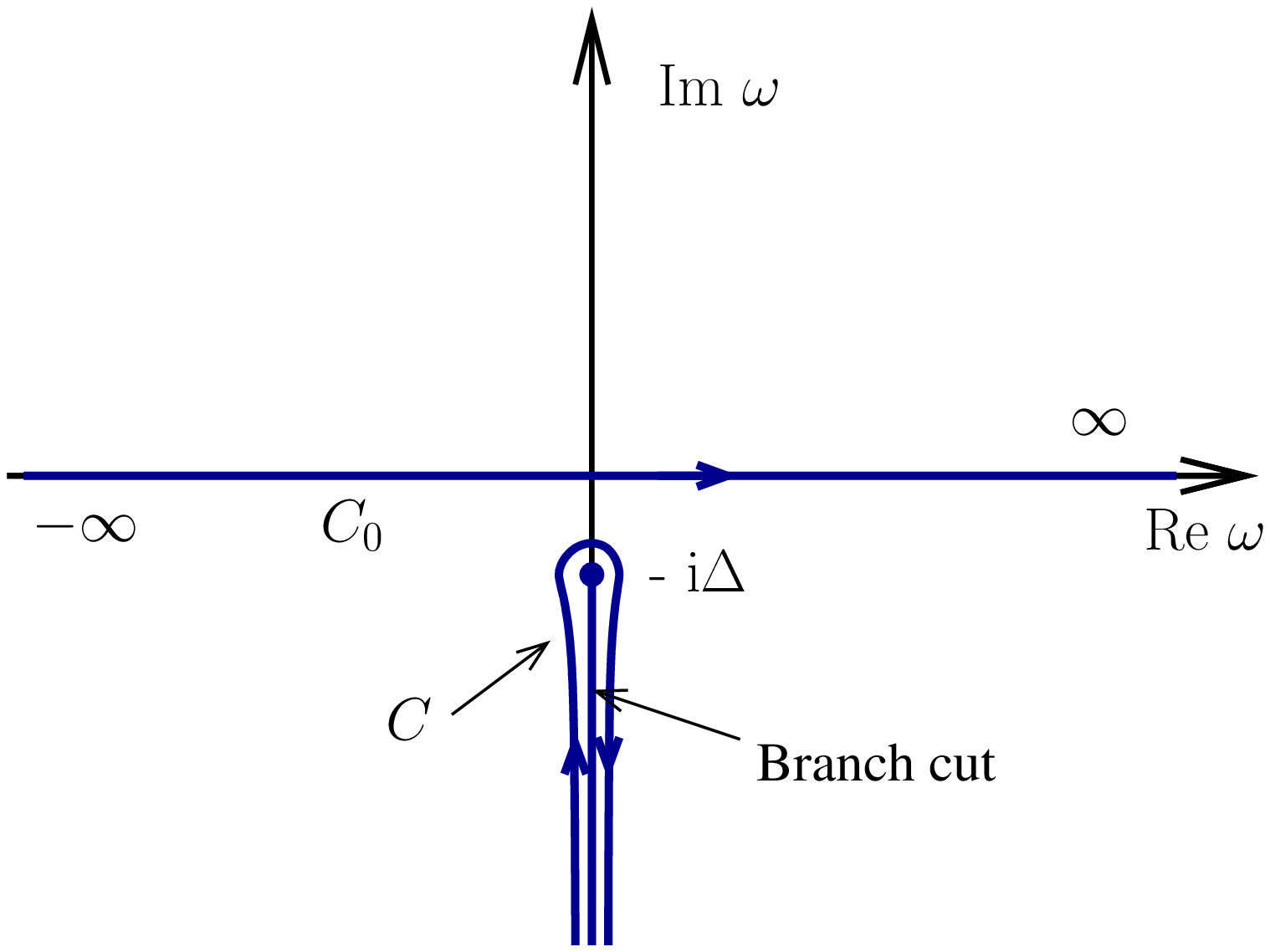}
 \includegraphics[width=0.95\columnwidth]{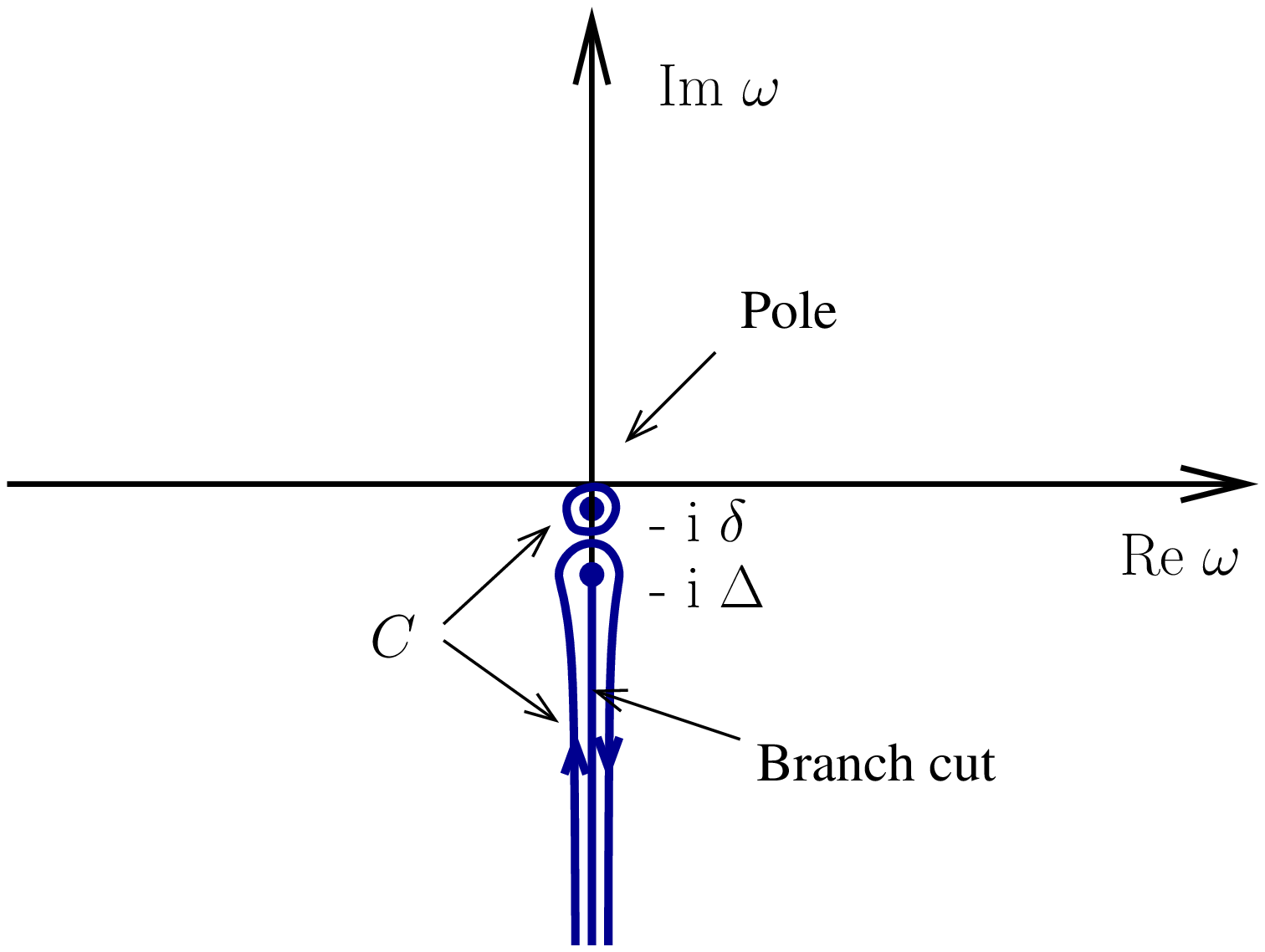}
  \caption{\label{fig:contour}
  (Color online). Pole structure of the integrand in Eq.~\eqref{eq:current_time}, 
for the spin-up -- spin-down (upper panel) and 
spin-up -- spin-up (lower panel) channels. 
In the spin-$\uparrow\downarrow$ channel, the pole at $\omega = 0$ 
is canceled by the $\omega^2$ dependence of $ g(\omega, T)$, while in the case of 
spin-$\uparrow\uparrow$ this pole survives and gives a finite response
as $t\to\infty$.}
\end{figure}

Surprisingly, just using Eq.~\eqref{eq:current_time} and the analytical 
properties of the functions  $ G_{LR}^{\s\uparrow} (\omega)$,
we are able to make rather strong  statements on the transient  
response, $ \expect{ J_{L}^{\s}(t)}$. Let us start by briefly summarizing these 
analytical properties. First of all, being retarded response 
functions, $  G_{LR}^{\s\s'} (\omega)$, are analytical on 
the upper half plane. Moreover, as assured by Fermi liquid 
theory, they are also analytical in an extended  region around 
$\w=0$  at any temperature. However, from perturbation theory 
we know that at very large frequencies, 
$\w\gg T_K,T$, they have  logarithmic  tails, 
$|G_{LR}^{\s\s'} (\omega)|\sim 1/\ln^2(\w/T_K)$ and thus tend to zero even 
in the universal scaling limit, $D\to\infty$, $T_K$~finite. Their 
asymptotic behavior and their symmetries
[$\myRe G_{rr'}^{\s\s'} (\omega) = \myRe G_{rr'}^{\s\s'} (-\omega)$ while $\Im m\; G_{rr'}^{\s\s'} (\omega) = 
-\Im m \;  G_{rr'}^{\s\s'} (-\omega)$] imply the presence of a {\em cut}
along the negative imaginary axis with an endpoint, $-i \Delta$,
with $\Delta\propto \max\{T,T_K\}$
(see Fig.~\ref{fig:contour}).~\footnote{We remark that while perturbation 
theory indeed reproduces this cut, the master equation approach 
fails to do that, and only produces a pole at $-i E_K$.}
Furthermore, as already discussed, 
$G_{rr'}^{\uparrow\downarrow}(\w=0) = 0$, while the components
$G_{rr'}^{\uparrow\uparrow  }(\w=0) = G_{rr'}(\w=0)/2$ remain finite.

Let us now discuss the properties of the response,  
Eq.~\eqref{eq:current_time}. First, we notice that 
due to the asymptotic $1/\ln^2(\w/T_K)$ fall-off of 
$G_{LR}^{\s\s'} (\omega)$ and the analyticity on the upper half-plane,
the integral contours in Eq.~\eqref{eq:current_time} can be closed 
upwards for any time $t\le 0$. Therefore, 
$\expect{ J_{L}^{\s}(t)}=0$ for $t\le 0$, i.e., it respects causality.
The response being zero 
even at $t=0$ is not entirely trivial: in the master equation approach, 
e.g., $G_{rr'}^{\uparrow\uparrow}(\w)$ remains finite 
in the $\omega\to\infty$ limit, and one obtains an unphysical 
 jump at $t=0$. We notice that 
the statement 
on the $t=0$ response being zero is equivalent to the  
Kramers-Kronig relation. Though the response 
$\expect{ J_{L}^{\s}(t)}$ vanishes at time $t=0$ and is continuous for 
times $t\ge 0$, the {\em slope} of the response, 
$\frac {\rm d}{{\rm d}t}\expect{ J_{L}^{\s}(t)}|_{t=0}$ is, however,
infinite, since the integral 
$\int_{-\infty}^\infty {\rm d}\w\; G_{rr'}^{\uparrow\uparrow}(\w)$ logarithmically diverges.  

For times $t>0$, the contours must be closed 
downwards, as shown in   Fig.~\ref{fig:contour}. In the spin-up -- spin-down
channel, $G_{LR}^{\uparrow\downarrow}(\w=0)=0$, 
and therefore the pole at $-i\delta$ does not give any 
contribution. The contribution of the cut to the 
spin up-down response can be written as
\begin{eqnarray}
 \frac {\expect{ J_{L}^{\downarrow}(t)}}{\delta V_{R}^{\uparrow}}
 &=& -  \int_{C} G_{LR}^{\uparrow\downarrow}(z)\; \frac{1}{z}\;
{\rm e}^{-i\,z\,t}\;{\rm d}z \,
\label{eq:contour}
\\
& = & 
{ \frac {e^2}{h} } 
\sin^2\phi\;
e^{-\Delta\, t}\int_{0}^{\infty}  \delta g(\Delta+y)
{\rm e}^{-y\, t} \; \frac {{\rm d}y}{y},
\nonumber
\end{eqnarray}
with $\delta g(x) =  2 \,\Im m\; g(-i \,x +\delta)$ 
the cut of the universal conductance function, $g(\w)$.
Clearly, the contribution of the cut falls off
as $\sim e^{-\Delta\,t}$ for long times.  
At $T=0$ temperature we have $\Delta\propto T_K$, and 
furthermore   $\delta g$ must be also a universal function, 
$\delta g(\Delta+y)= \delta g(y/T_K)$. 
Therefore, the response is a universal function 
of $t\, T_K$. We can also tell the short time asymptotics 
of the response. Making use of the fact that
the response is continuous at $t=0$, we obtain for 
$t\ll 1/\Delta$ the expression, 
\be
 {\expect{ J_{L}^{\downarrow}(t)}}
 \sim  
{ \frac {e^2}{h}   }
\sin^2\phi\;
\int_{0}^{\infty}  \delta g(\Delta+y)
({\rm e}^{-y\, t}-1) \; \frac {{\rm d}y}{y}.
\ee
Since the cut scales for large energies as 
$\sim 1/\ln^3(y/T_K)$, we get,
\be
{\expect{ 
{ J_{L}^{\downarrow} }
(t\ll 1/\Delta)}}
\sim \frac{\delta V_{R}^{\uparrow}}{\ln^2\frac 1{t\,T_K}}\;\;.
\label{logsing}
\ee
Remarkably, this result does not depend on the temperature, since 
it is determined only by the high frequency part of the conductance. 
Furthermore, since the length of the current pulse is determined by
the exponential prefactor, $\sim{\rm e}^{-\Delta\;t}$, we can also 
read out of Eq.~\eqref{logsing} its  height: 
for  $T\ll T_K$ one has $\Delta\sim T_K$, 
and the height of the pulse is 
${\expect{ J_{L}^{\downarrow}(t)}} \sim \delta V_{R}^{\uparrow}$. 
For $T\gg T_K$, on the other hand, we have $\Delta\sim T$, 
and the height of the current pulse, is $\sim {\delta V_{R}^{\uparrow}}/{\ln^2\frac T{T_K}}$. 

\begin{figure}[t]
  \includegraphics[width=0.95\columnwidth]{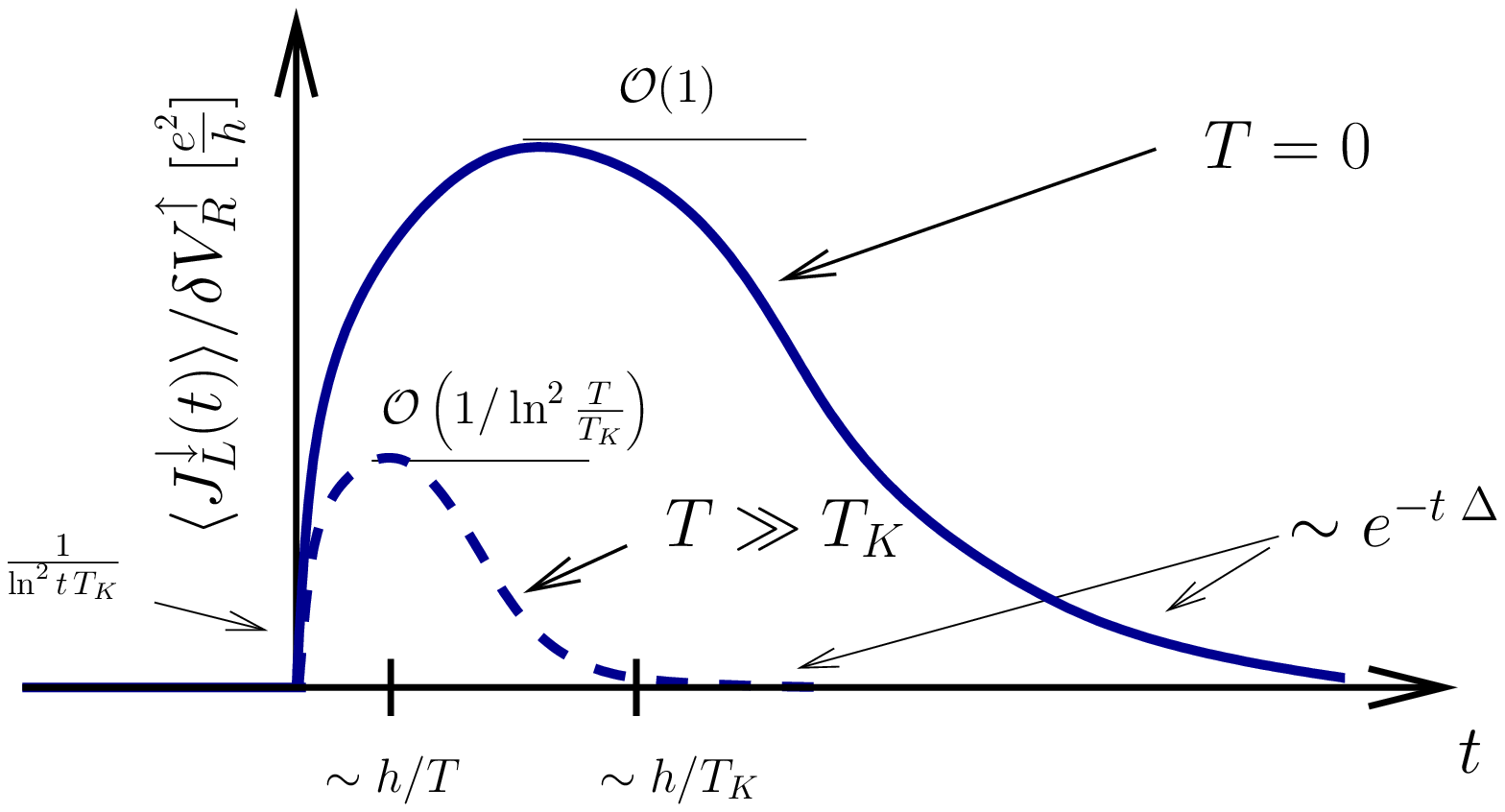}
  \includegraphics[width=0.95\columnwidth]{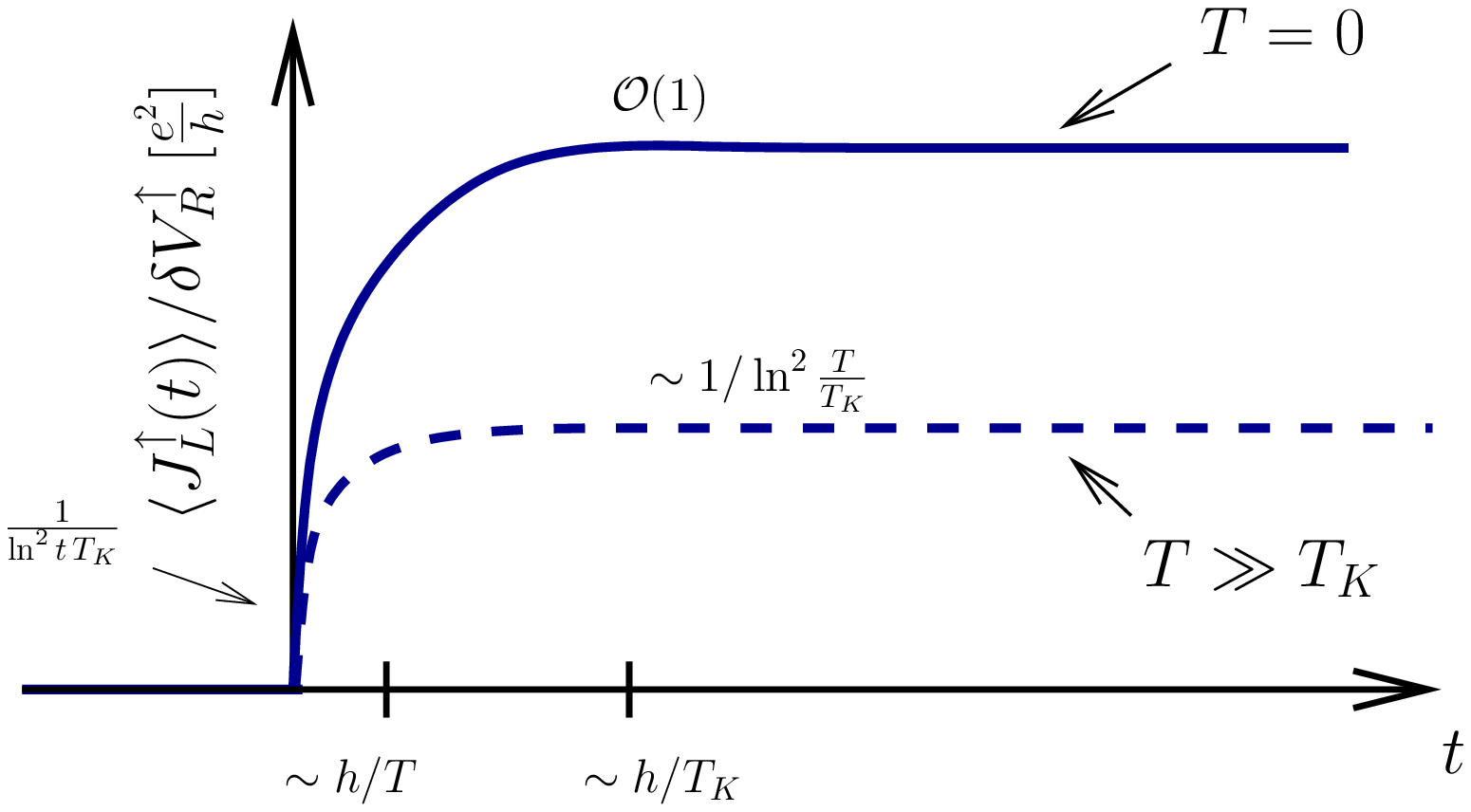}
  \caption{\label{fig:current}
  (Color online). Transient current response in the spin-down (upper panel)
 and spin-up (lower panel) channels, upon a constant bias 
 applied at ${\rm t}=0$ in the spin-up channel. 
}
\end{figure}

Figure \ref{fig:current} summarizes  all the 
 characteristic features of the current response 
$\expect{ J_{L}^{\downarrow}(t)}$, discussed above. 
The total charge pumped into spin-down channel of left lead
is simply given by the integral of the transient 
response, and is approximately
\be
\Delta Q_{L}^{\downarrow} \sim 
{ \frac {e^2}{h} }
\sin^2\phi 
\begin{cases}\frac{1}{T\ln^2(T/T_K)}  , & \mbox{if} \;\; T\ll T_K\;, \\ 
\frac{1}{T_K}, 
 & \mbox{if} \;\; T\gg T_K \;.
\end{cases}
\ee
Remarkably, the coefficients appearing in this expression are 
just the high-temperature and low temperature expressions of the 
spin susceptibility.~\cite{BetheAnsatz}

The analysis of the response of the spin-up carriers 
follows very similar lines. The only major difference is that 
in this case the pole at $-i\delta$ gives a finite time independent 
contribution, and  leads to an asymptotic response, 
\be
{\expect{ 
{ J_{L}^{\uparrow}(t\to\infty)} }
} = G_{LR}^{\uparrow\uparrow}(\w=0)\;
\delta V_{R}^{\uparrow}\;.
\ee
Otherwise, our discussions on the universal form of the response, and 
its short time $1/\ln^2(t)$ singularity carry over to this case, too. 
Instead of giving further 
details on  ${\expect{ J_{L}^{\uparrow}(t\to\infty)}} $, 
we just summarized its properties in Fig.~\ref{fig:current}.


\section{Spin relaxation effects} 
\label{sec:spin_relaxation}

All results presented so far were obtained under the assumption that
spin relaxation is generated by the exchange coupling $j$, and 
there are no external sources of spin relaxation. 
In reality, however, external spin relaxation channels are 
always present. In quantum dots, the dominant channel of 
(external) spin relaxation is usually due to 
hyperfine interaction between the confined electron and 
nuclear spins in the host material, leading  
typically to a dephasing time of the order 
of $\tau_s\sim 10{\; \rm ns}$ or longer in the absence of 
magnetic field.~\cite{LossGlazman,LairdMarcus}
These hyperfine relaxation 
processes are thus characterized 
by an energy scale $h /\tau_s\sim 1-10 {\;\rm mK}$, 
typically much smaller than the temperature.
Coupling to piezoelectric phonons~\cite{KaetskiNazarov} 
or electromagnetic fluctuations~\cite{SanJose} 
through spin-orbit interaction or polaron dephasing processes due to coherent 
acoustic phonon generation are,
in general, characterized by even longer dephasing times and smaller relaxation rates.~\cite{Zumbuhl} 
Therefore, for typical experimental parameters, we would naively 
expect  $1/\tau_s$ to be small compared to the temperature, $T$. 
Nevertheless, a finite $\tau_s$
 leads to qualitatively different results, since 
its presence lifts the constraint of spin conservation, and allows to 
have a finite d.c. spin cross-conductance, 
$G_{LR}^{\uparrow\downarrow}(\w=0)\ne 0$.

\begin{figure}
  \includegraphics[width=0.95\columnwidth]{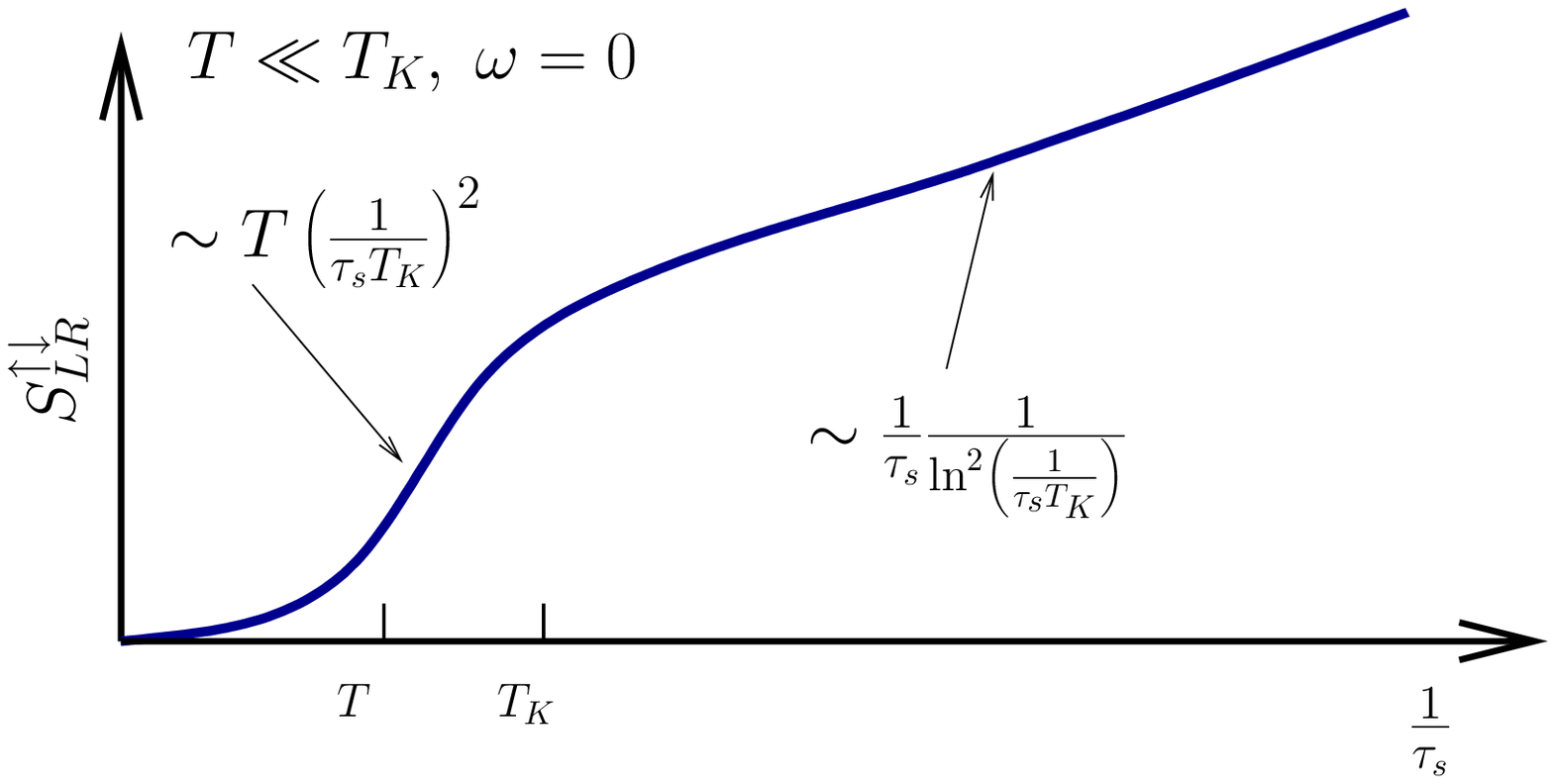}
  \includegraphics[width=0.95\columnwidth]{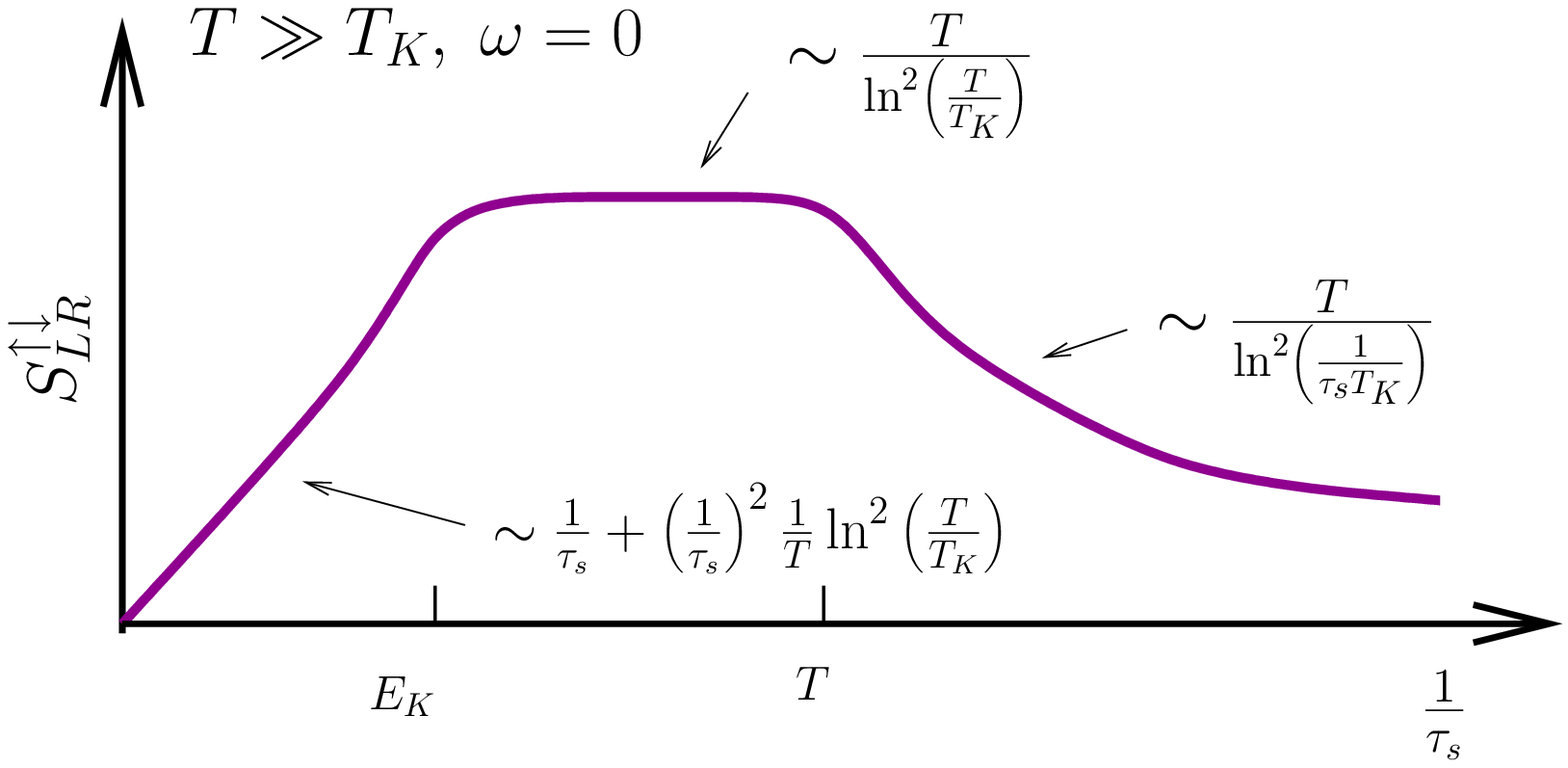}
  \caption{\label{fig:S_tau_s}
  (Color online)   
Shot noise $S_{LR}^{\uparrow\downarrow}(\omega=0, T) $
as a function of the spin relaxation rate $1\over \tau_s$ in the Fermi liquid
regime ($T\ll T_K$) (upper panel) 
and in the perturbative regime ($T\gg T_k$) (lower panel).}
\end{figure}

To investigate the effect of a finite $\tau_s$, let us consider the
perturbative ($T\gg T_K$) and Fermi liquid ($T\ll T_K$) regimes 
separately. In the regime $T\gg T_K$, we can readily extend our master equation
analysis to include $1/\tau_s$ and obtain,   
\begin{equation}
   S_{LR}^{\uparrow \downarrow} (\omega<T)\simeq
  -
{\frac{e^2}{h} }
\;\frac{\pi}{4}  \frac{E_K\; 
\left (\omega^2+\frac 1  {\tau_s^2} +\frac{E_K}{\tau_s} \right)
  \sin^2 \phi} {\omega^2+E_K+\frac {1}{\tau_s^2}}\;\nonumber.
\end{equation}
Here we incorporated already logarithmic corrections by replacing 
the bare Korringa rate, $E_{K,0}$ by its renormalized value, 
given by 
Eq.~\eqref{eq:Korringa}. Clearly, at large frequencies, 
$\w\gg E_K$, external spin relaxation does not play a role.
However, for $\w< E_K$, it
suppresses the dip in  $S_{LR}^{\uparrow \downarrow}(\w)$, and 
leads to a finite cross-spin "shot noise",
\bea
   S_{LR}^{\uparrow \downarrow} (\w=0)&\approx&
  -
{\frac{e^2}{h} }\;  \sin^2 \phi
\;\frac{\pi}{4}  
\frac{E_K\;\frac 1 {\tau_s}}{E_K + \frac 1 {\tau_s}} \;
\nonumber 
\\
&\approx& 
-\sin^2 \phi
{  \frac{e^2}{h}   }
\;\frac{\pi}{4}\;
\min\{E_K , \frac 1 {\tau_s} \}\;.
\eea
In other words, in the most relevant case, 
$ 1/ {\tau_s}< E_K$, the cross-spin current noise is 
simply proportional to $1/\tau_s$, as also found by
Kindermann.~\cite{KindermannPRB2005}
The full frequency spectrum of the noise 
in the perturbative regime ($T\gg T_K$)
is presented in Fig.~\ref{fig:shot_noise_finite_T_comparison}, 
lower panel.

In the Fermi liquid regime, $T \ll T_K$, finite external spin relaxation
also leads to a finite cross-spin noise. To estimate it, we first
notice that $1/\tau_s$ just introduces a new frequency scale, and therefore we
expect
\be 
  G_{LR}^{\uparrow \downarrow} (\omega\to 0)\sim 
 \frac{1}{\tau_s^2 T_K^2}\;.
\ee 
Notice that $T$ does not appear in this equation. 
Correspondingly, the noise behaves for $T< T_K$ as
\be 
S_{LR}^{\uparrow \downarrow} (\omega\to 0)\sim 
\frac{T}{\tau_s^2 T_K^2}\;.
\ee

The overall dependence of the cross-spin shot noise signal 
on the spin relaxation rate is sketched in Fig.~\ref{fig:S_tau_s}. There we also display the 
physically not too relevant regime, $1/\tau_s> T$, where
 $1/\tau_s$ becomes the dominant energy scale, 
and therefore we have $j\to  1/\ln(1/\tau_s T_K)$.


\section{Conclusions}

In the present paper we studied the equilibrium 
spin current noise and spin conductance through a quantum dot in its 
Kondo (local moment) regime. We have shown that in the absence of 
external fields, they are both characterized by a pair of universal 
functions, and determined the properties of these functions. 
We have shown that 
 -- in contrast to the charge conductance ($G_{LR}$) -- 
 the d.c. spin cross-conductance  ($G^{\uparrow \downarrow}_{LR}$) 
vanishes. Put in another way, there is no spin drag, and a spin-up current 
cannot generate a steady spin-down current, at least not 
within linear response. 
At $T=0$ temperature this obviously follows from   Fermi liquid properties, 
and is thus valid for {\em any} interacting 
system with no spin-orbit coupling and with a Fermi liquid ground state. 
However, somewhat surprisingly, though it is not true for  any interacting
system, for a quantum dot, this property also carries over  for finite temperatures. 
It is related to the simple structure of the Kondo (or the underlying Anderson)  
models, where spin transfer between spin-up and spin-down states 
can occur only through a single point, namely the dot state
(or the dot spin in the Kondo model). Therefore, 
the spin currents generated by consecutive $\Uparrow\to\Downarrow$ 
and $\Downarrow\to\Uparrow$  flips of the dot spin precisely 
cancel each-other, and no d.c. cross-spin currents appear. 
Correspondingly, the cross-spin shot noise, $S^ {\uparrow\downarrow}_{LR}(\w=0,T)$ 
also vanishes at {\em any} temperature, and the  noise spectra, 
 $S^ {\uparrow\downarrow}_{LR}(\w)$ and  $S^ {\uparrow\uparrow}_{LR}(\w)$
both exhibit  related low frequency anomalies. 

External spin relaxation slightly changes the picture above. 
It partly removes the correlations between consecutive 
spin-flip processes, and makes $G^ {\uparrow\downarrow}_{LR}(\w=0)$
and $S^ {\uparrow\downarrow}_{LR}(\w=0)$ finite. However, 
since the external spin-flip rate, $1/\tau_s$ is typically
much smaller than the other energy scales ($T$, $T_K$, $E_K$), 
it only leads to  small changes in the overall behavior of the 
noise and conductance functions.

As we also demonstrated in detail, simple-minded perturbation theory 
accounts only for the structure of individual coherent processes, and 
fails badly to capture these correlations between consecutive 
processes, which happen to dominate the spin response at small 
frequencies. Therefore, one must be very careful when calculating spin transport properties.
Even in the perturbative regime, $T\gg T_K$, simple-minded perturbation 
theory is valid only for frequencies above the Korringa rate, 
$\w> E_K$. To capture the physics at frequencies
$\w<E_K$, a supplementary master equation approach (valid for $\w< T$) 
can be employed. Alternatively, one can use 
a more systematic but also  more technical quantum Langevin approach, 
which works for any frequency in the perturbative regime, 
$T\gg T_K$, but neglects logarithmic corrections (see Ref.~[\onlinecite{future},\onlinecite{Konig1996}]). 

\begin{figure}[t]
  \includegraphics[width=0.98\columnwidth]{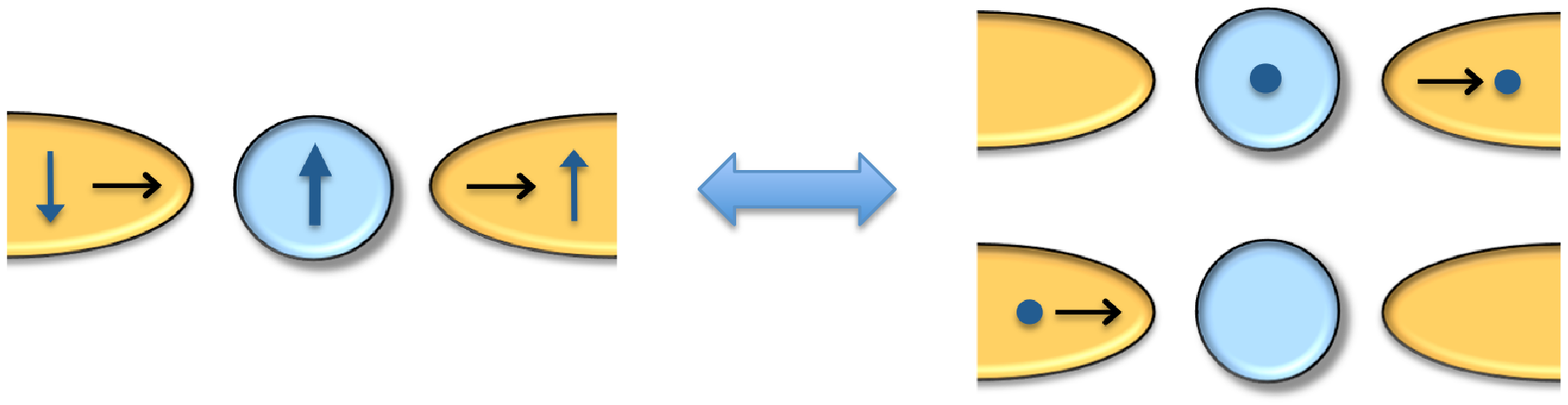}
  \caption{\label{fig:DD_mapping}
  (Color online) Correspondence between a single quantum dot device
  and a spinless double dot device.}  
\end{figure}
 
Although finite-frequency noise measurements
are now available,~\cite{Clarke,Heiblum,Deblock03, OnacPRL2006_2, reulet_PRL08, delattre_NatPhys09}
and spin polarized currents can also be relatively 
easily produced,\cite{Folk_PRL09,Kontos}
measuring the low-frequency anomalies
of spin cross-correlations, $S^ {\uparrow\downarrow}_{LR}(\w)$,
seems to be a difficult task.
However, the predicted low frequency anomalies 
are also  present in the spin polarized conductance, 
$G_{LR}^{\uparrow\uparrow}(\w)$, and noise,
 $S^ {\uparrow\uparrow}_{LR}(\w)$ (see Figs.~\ref{fig:Gupup_simple} and 
\ref{fig:s_simple}). These are experimentally much more easily accessible, since 
carriers must be polarized in the same direction. 
Alternatively, one can measure these cross-correlations in the 
{\em charge sector}, by using capacitively coupled double dots
(see Fig.~\ref{fig:DD_mapping}).\cite{Hettler,McClure} 
In the spin polarized case,  
the Hamiltonian of the double dot system 
maps to that  of the Anderson model with anisotropic 
hybridization parameters. Measuring 
noise or conductance between leads attached to the upper or lower 
leads of the double dot device 
shown in Fig.~\ref{fig:DD_mapping} is thus equivalent 
to cross-spin measurements in the single quantum dot setup.

\begin{acknowledgments}
This research has been supported by Hungarian Scientific
Research Funds Nos.  K73361, CNK80991,
T\'{A}MOP-4.2.1/B-09/1/KMR-2010-0002, the 
Romanian grant CNCSIS PN II ID-672/2008, and
the EU-NKTH GEOMDISS project.
I. W. acknowledges support from
the Ministry of Science and Higher Education through a
research project "Iuventus Plus" in years 2010-2011 and
the Alexander von Humboldt Foundation.
\end{acknowledgments}


\end{document}